\documentclass{JFM-FLM_Au}

\lefttitle{A. Rehman and D. Floryan}
\righttitle{Journal of Fluid Mechanics}

\title{Propulsion of a flexible foil in a wavy flow: resonance, antiresonance, and destructive self-interference}

\author{Abdur Rehman\aff{1}, Daniel Floryan\aff{1} }

\affiliation{\aff{1}Department of Mechanical and Aerospace Engineering, University of Houston, Houston, TX 77204, USA}

\corresau{Abdur Rehman, \email{arehman12@uh.edu}; Daniel Floryan, \email{dfloryan@uh.edu}}

\begin{document}
\maketitle

\begin{abstract}
Swimming and flying animals demonstrate remarkable adaptations to diverse flow conditions in their environments. In this study, we aim to advance the fundamental understanding of the interaction between flexible bodies and heterogeneous flow conditions. We develop a linear inviscid model of an elastically mounted foil that passively pitches in response to a prescribed heaving motion and an incoming flow that consists of a traveling wave disturbance superposed on a uniform flow. In addition to the well-known resonant response, the wavy flow induces an antiresonant response for non-dimensional phase velocities near unity due to the emergence of non-circulatory forces that oppose circulatory forces. We also find that the wavy flow destructively interferes with itself, effectively rendering the foil a low-pass filter. The net result is that the waviness of the flow always improves thrust and efficiency when the wavy flow is of a different frequency than the prescribed heaving motion. Such a simple statement cannot be made when the wavy flow and heaving motion have the same frequency. Depending on the wavenumber and relative phase, the two may work in concert or in opposition, but they do open the possibility of simultaneous propulsion and net energy extraction from the flow, which, according to our model, is impossible in a uniform flow. 
\end{abstract}


\section{\label{sec:level1}Introduction}

 Natural fliers and swimmers, like birds and fish, efficiently navigate complex environments with their flexible bodies. In contrast, man-made flying and swimming machines may struggle in complex environments they were not designed for, leading to increased energy consumption. Aiming to replicate nature's efficiency, improve propulsive performance, and reduce the acoustic signatures of flying and swimming machines, biologists and engineers have turned their attention towards bio-inspired locomotion \citep{liao2007review, smits2019undulatory}. 

 Much of our understanding of bio-inspired locomotion is based on the study of pitching and heaving foils \citep{smits2019undulatory}, which serve as models of the wings and fins of flying and swimming animals \citep{wu2011fish}. The theoretical foundations for flapping foils were developed in the context of the aerodynamic flutter of wings. \cite{theodorsen1935general} and \cite{garrick1936propulsion} developed the unsteady, inviscid, linear, thin airfoil theory for a foil undergoing oscillatory pitching and heaving motions in the presence of a uniform flow, deriving expressions for the forces and moments on the foil. Recently, there has been some debate about the linear theory \citep{fernandez2016linearized, gordillo2025note}. \Citet{von1938airfoil} developed the thin airfoil theory for a static foil encountering a varying vertical velocity due to a stationary gust. \cite{wu1972extraction} combined and extended these two theories to develop the linear theory for a flapping foil encountering a periodically varying transverse flow induced by waves, his interest being in extracting energy from waves. Later studies \citep{goldstein1976complete, massaro2015effect, lysak2013prediction, taha2019viscous, catlett2020unsteady} explored higher-order effects arising due to three-dimensional flow, viscosity, foil thickness, and finite flapping amplitudes.
 
Beyond the foundational theoretical work, a considerable amount of research has been conducted on flapping foil propulsion, primarily in a uniform flow. Studies have explored a range of factors contributing to propulsive performance, including kinematics \citep{anderson1998oscillating, young2007mechanisms, floryan2017scaling, van2019scaling, floryan2019large}, the role of wake vortices generated behind the flapping foil \citep{smits2019undulatory, zhang2017footprints, floryan2020swimmers}, flexibility in chordwise \citep{spagnolie2010surprising, moore2015torsional} and spanwise directions \citep{liu1997propulsive, heathcote2008effect, gordnier2013high}, the aspect ratio of the foil \citep{chopra1974hydromechanics, buchholz2008wake, dewey2013scaling, ayancik2019scaling}, the role of fluid-structure resonance \citep{alben2008optimal, michelin2009resonance, dewey2013scaling, quinn2014scaling, paraz2016thrust, floryan2018clarifying, goza2020connections}, and optimal Strouhal numbers \citep{triantafyllou1991wake, taylor2003flying, gazzola2014scaling, floryan2018efficient}. This extensive knowledge has led to flying and swimming machines that are often designed and optimized for uniform flows.

Animals, on the other hand, skillfully adapt to their complex and unsteady flow environments. Animals' adaptations not only mitigate potentially deleterious effects of their flow environments, but also allow them to benefit from complex flows. The most familiar example is thermal soaring, where birds utilize rising hot air columns to fly long distances with minimal flapping effort and energy expenditure \citep{newton2024migration}. In another example, it was demonstrated that a freshly killed trout can hold station in the presence of an incoming flow---that is, a dead fish can swim---by utilizing passive deformations of its flexible body to extract energy from incoming unsteady vortical flow \citep{beal2006passive}. The first example demonstrates a behavioral adaptation, while the second demonstrates an adaptation in the physical design that allows the animal to take advantage of its complex flow environment. 

Most studies on locomotion in complex and unsteady flow environments are motivated by the collective motion of a group of animals, where individuals must contend with flow alterations due to the presence and actions of other individuals in the group (although there are other settings that have been studied as well \citep{liao2007review}). It has been argued that the wakes of swimmers provide opportunities for following swimmers to save energy compared to swimming in isolation \citep{weihs1973hydromechanics}. Studies range from extensions of the classical unsteady thin airfoil theory to multiple interacting foils \citep{baddoo2023generalization} to applications of reinforcement learning to learn efficient locomotion strategies in the wakes of other swimmers and obstacles \citep{verma2018efficient, zhu2021numerical}. Insights from schooling swimmers have even been used to develop configurations of arrays of vertical axis wind turbines that exploit each other's wakes for enhanced energy extraction \citep{whittlesey2010fish}.

Although locomotion in complex and unsteady flow environments has been studied under specific contexts, a comprehensive understanding eludes us. Moreover, flexibility and the attendant fluid-structure interaction seem to be key physical enablers that allow one to exploit a complex flow environment (as in the example of the dead fish that is able to swim \citep{beal2006passive}), but the effects of flexibility in complex and unsteady flows are not systematically understood. 
 
 In this work, we seek a fundamental understanding of how animals may use their flexible bodies to exploit spatially and temporally varying flows for enhanced locomotion. This builds on prior work that developed an understanding of how animals may exploit spatial and temporal heterogeneity of their bodies \citep{floryan2020distributed, yudin2023propulsive}. To thoroughly explore this phenomenon, we use an analytical approach, which allows for an in-depth investigation while avoiding computational and experimental expenses. Our approach uses the linear, inviscid, thin airfoil theory of \cite{wu1972extraction} to model the fluid mechanics and the approach of \cite{moore2014analytical} to model flexibility. The flow consists of a uniform flow with a superposed harmonic (in space and time) traveling wave disturbance in the velocity field. Such a description of the flow is generic since any disturbance can be expanded into a superposition of traveling waves. Our primary goal is to elucidate the fundamental physical principles governing flapping propulsion in heterogeneous flows. 

 After describing the problem and mathematical model in Section~\ref{sec:prob}, we study the emergent kinematics in response to the wavy flow and active motion in Section~\ref{subsec:freq_dom}. The resulting propulsive performance is tackled in Sections~\ref{sec:unequal} and~\ref{sec:equal_sigma}, the former addressing the case where the wavy flow and active motion have different frequencies, and the latter the case where the two have equal frequencies. A summary and concluding remarks are presented in Section~\ref{sec:conclusion}.

\section{Problem description and mathematical modeling}
\label{sec:prob}

\begin{figure}
  \centerline{\includegraphics[width=0.5\linewidth]{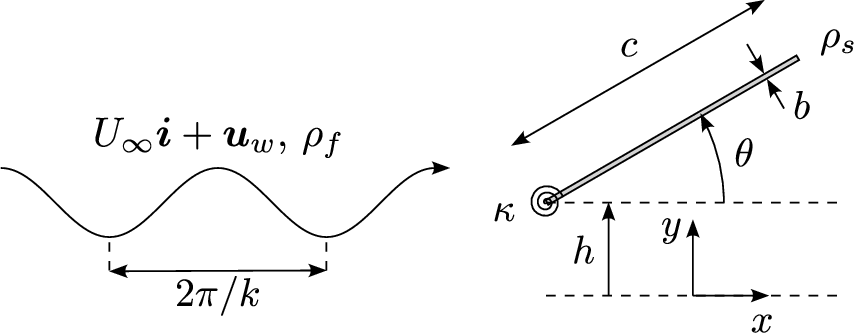}}% Images in 100% size
  \caption{Schematic of the problem}
\label{fig:setup}
\end{figure}

Consider a thin foil of chord $c$, thickness $b$, span $s$, and density $\rho_s$, as sketched in figure~\ref{fig:setup}. The foil is driven at its leading edge by a sinusoidal heaving motion $h$ of amplitude $h_0$ and frequency $f$, and is held against a freestream velocity consisting of a uniform flow with speed $U_\infty$ and a wavy flow $\boldsymbol{u}_w$, with $\vert \boldsymbol{u}_w \vert \ll U_\infty$. At $y = 0$, coincident with the mean position of the foil, the horizontal component of the wavy flow is $U_w \textrm{e}^{\textrm{i}(2\upi f_w t - kx)}$ and the vertical component is $\textrm{i}V_w \textrm{e}^{\textrm{i}(2\upi f_w t - kx)}$, with $\textrm{i} = \sqrt{-1}$. That is, the wavy flow takes the form of a traveling wave with encounter frequency $f_w$ and wavenumber $k$. We use complex numbers for later convenience; here and throughout, the real parts of complex expressions should be taken when evaluating physical variables. Without loss of generality, $V_w \in \mathbb{R}$. At the leading edge, a torsional spring of stiffness $\kappa$ allows the foil to pitch passively with pitch angle $\theta$ in response to the heaving motion and wavy flow.

The fluid is Newtonian, inviscid, and has a constant density $\rho_f$. Assuming $h_0 \ll c$, at leading order, conservation of mass and momentum for the fluid lead to
 \begin{align}
    \boldsymbol{\nabla \cdot u} &= 0, \label{eq:prob3}\\
    \rho_f (\boldsymbol{u}_t + U_\infty\boldsymbol{u}_x) &= -\boldsymbol{\nabla}p, \label{eq:prob4}
 \end{align}
 where $\boldsymbol{u} = u\boldsymbol{i} + v\boldsymbol{j}$ is the perturbation velocity induced by the relative motion between the fluid and the foil, $\boldsymbol{i}$ is the unit vector in the $x$-direction, $\boldsymbol{j}$ is the unit vector in the $y$-direction, $p$ is the associated perturbation pressure, subscript $t$ denotes differentiation with respect to time, and subscript $x$ denotes differentiation with respect to the streamwise coordinate. Identical equations hold for the wavy velocity and pressure fields. The perturbation velocity additionally satisfies the no-penetration boundary condition on the surface of the foil, it decays to zero in the far field, and it is subject to the Kutta condition at the trailing edge of the foil; see~\cite{wu1972extraction} for details. 

 A few important features are worth noting. The first is that the application of the Kutta condition at the trailing edge, together with Kelvin's circulation theorem, implies that vorticity must be shed from the trailing edge. This takes the form of a thin vortex sheet attached to the trailing edge of the foil. The vortex sheet is assumed to be planar (consistent with the small-amplitude assumption), and it extends downstream to infinity. Second, at leading order, the total velocity and pressure fields are superpositions of those due to the uniform flow, the wavy flow, and the presence of the foil. Furthermore, the wavy flow and the flow due to the presence of the foil are decoupled in the governing equations. The wavy velocity field does, however, appear in the no-penetration boundary condition on the surface of the foil, 
\begin{equation}
  \label{eq:bc}
  v|_{y = 0} = \dot h - v_w|_{y = 0} + \dot \theta x + U_{\infty} \theta,
\end{equation}
where $v_w = \boldsymbol{u}_w \boldsymbol{\cdot} \boldsymbol{j}$. It is solely through this boundary condition that the wavy velocity plays a part in determining the velocity and pressure fields associated with the presence of the foil \citep{wu1972extraction}. Note that only the vertical component of the wavy velocity field enters at leading order. For $\vert \boldsymbol{u}_w \vert \ll U_\infty$ and $h_0 \ll c$, the effects of the horizontal component of the wavy velocity are therefore negligible. Moreover, only the wavy velocity field along $y = 0$ appears in the leading-order boundary condition. This is to say that, at the leading order of approximation, the horizontal component of the wavy velocity field plays no part in determining the dynamics and forces of a flapping foil, nor does the vertical structure of the wavy flow.

\subsection{Fluid-structure interaction}

 The passive pitching motion of the foil is driven by the spring, inertial, and fluid moments. An angular momentum balance yields the governing dynamics for the pitching motion,
\begin{equation}
    I \ddot\theta = -\kappa \theta -\frac{\rho_s bc^2 s \ddot{h}}{2} + M_f,
    \label{eq:prob1}
 \end{equation}
 where $I = \frac{1}{3}\rho_s bc^3 s$ is the second moment of inertia about the leading edge, the second term on the right-hand side is the inertial moment about the leading edge that arises from the vertical acceleration of the leading edge, and 
 \begin{equation}    
    M_f = \int_{-c/2}^{c/2} \Delta p \left(x + \frac{c}{2}\right) s \mathrm{d} x
    \label{eq:prob2}
 \end{equation}
 is the moment about the leading edge induced by the fluid, where $\Delta p$ is the pressure difference across the foil. Throughout, we fix the leading edge of the foil at $x = -\frac{c}{2}$. Note that the pressure field associated with the wavy flow does not induce any force or moment on the thin foil since it is continuous across the foil. In particular, the wavy pressure field does not contribute to the moment $M_f$ in~\eqref{eq:prob2}. 

 To non-dimensionalize the problem, we use $\frac{c}{2}$ as the length scale, $U_\infty$ as the velocity scale, and $\frac{c}{2U_\infty}$ as the time scale. The non-dimensional form of the angular momentum balance is
 \begin{equation}
    \label{eq:prob5}
    \frac{16}{3}R\ddot{\theta} = -4K\theta - 4R\ddot{h}^* + C_M,
 \end{equation}
 where
 \begin{equation}
    \label{eq:prob6}
    R = \frac{\rho_s b}{\rho_f c}, \quad K = \frac{\kappa}{\rho_f U_\infty^2 c^2 s}, \quad h^* = \frac{2h}{c}, \quad C_M = \frac{4M_f}{\rho_f U_\infty^2 c^2 s}.
 \end{equation}
 $R$ is the ratio of a characteristic mass of the foil to a characteristic mass of the fluid, and $K$ is the ratio of a characteristic moment from the torsional spring to a characteristic moment from the fluid. The corresponding scale for the pressure is $\rho_f U_\infty^2$. We call $R$ the mass ratio and $K$ the stiffness ratio. Henceforth, all variables are non-dimensional. 

 To solve for the unknown pitch angle $\theta$, we must obtain an expression for the moment coefficient $C_M$. The pressure difference across the foil can be written in terms of the displacement of the foil. Let the non-dimensional heaving motion take the form ${h^*(t) = h_0^* \textrm{e}^{\textrm{i}\sigma t}}$, where $\sigma = \frac{\upi fc}{U_\infty}$ is the reduced frequency, and let $\sigma_w = \frac{\upi f_w c}{U_\infty}$ be the reduced frequency of the wavy flow. Then the displacement of the foil is
 \begin{equation}
    \label{eq:prob7}
    Y(x,t) = h_0^* \textrm{e}^{\textrm{i} \sigma t} + (\theta_{0h} \textrm{e}^{\textrm{i} \sigma t} + \theta_{0w} \textrm{e}^{\textrm{i} \sigma_w t})(x + 1), \quad -1 \le x \le 1.
\end{equation}
Above, we have decomposed the passive pitching motion into motions due to heave and due to the wavy flow. Note that $h_0^*, \theta_{0h}, \theta_{0w} \in \mathbb{C}$, with their magnitudes and arguments giving the amplitudes and phases, respectively, of the corresponding motions. For later convenience, we rewrite the displacement as
 \begin{equation}
    \label{eq:prob8}
    Y(x, t) = \left(\frac{\beta_0}{2} + \beta_1 x\right)\textrm{e}^{\textrm{i} \sigma t} + \left(\frac{\beta_{0w}}{2} + \beta_{1w}x\right)\textrm{e}^{\textrm{i} \sigma_w t}, \quad -1 \le x \le 1,
 \end{equation}
 where $\beta_0 = 2h_0^* + 2\theta_{0h}$, $\beta_1 = \theta_{0h}$, $\beta_{0w} = 2\theta_{0w}$, and $\beta_{1w} = \theta_{0w}$. 

In terms of non-dimensional variables, the moment coefficient is equal to
 \begin{equation}
    \label{eq:kin1}
    C_M = \int_{-1}^1 \Delta p (x + 1) \mathrm{d}x.
 \end{equation}
 Following \citet{wu1972extraction}, for a known deflection of the form in~\eqref{eq:prob8}, the integral evaluates to
 \begin{equation}
    \label{eq:kin2}
    C_M = \textrm{e}^{\textrm{i} \sigma t}(\beta_0 a_0 + \beta_1 a_1) + \textrm{e}^{\textrm{i} \sigma_w t}(\beta_{0w}a_{0w} + \beta_{1w}a_{1w} + V_w^* a_w),
 \end{equation}
 where $V_w^* = \frac{V_w}{U_\infty}$ and the coefficients are
 \begin{align}
    \label{eq:kin3}
    a_0 &= -\frac{\upi \textrm{i} \sigma}{2} C(\textrm{i} \sigma) + \frac{\upi \sigma^2}{2}, \\
    a_1 &= -\upi C(\textrm{i} \sigma) - \frac{\upi \textrm{i} \sigma}{2} C(\textrm{i} \sigma) - \frac{3 \upi \textrm{i} \sigma}{2} + \frac{\upi \sigma^2}{8}, \\
    a_{0w} &= -\frac{\upi \textrm{i} \sigma_w}{2} C(\textrm{i} \sigma_w) + \frac{\upi \sigma_w^2}{2}, \\
    a_{1w} &= -\upi C(\textrm{i} \sigma_w) - \frac{\upi \textrm{i} \sigma_w}{2} C(\textrm{i} \sigma_w) - \frac{3 \upi \textrm{i} \sigma_w}{2} + \frac{\upi \sigma_w^2}{8}, \\
    a_w &= -\upi [W_1(\textrm{i} \sigma_w) - \textrm{i} W_2(\textrm{i} \sigma_w)] + 2 \upi \left( 1 - \frac{\sigma_w}{k^*} \right) J_1(k^*) - \upi \textrm{i} \left( 1 - \frac{\sigma_w}{k^*} \right) J_2(k^*).
 \end{align}
 Above, $k^* = \frac{kc}{2}$ is the non-dimensional wavenumber of the wavy flow, $J_n$ is the Bessel function of the first kind of order $n$, $C$ is the Theodorsen function, $F = \Real(C)$, $G = \Imag(C)$, $W_1 = (1 - F)J_1(k^*) + GJ_0(k^*)$, and $W_2 = FJ_0(k^*) + GJ_1(k^*)$. Theodorsen's function is defined by
 \begin{equation}
   C(s) = \frac{K_1(s)}{K_0(s) + K_1(s)},
 \end{equation}
 where $K_\nu$ is the modified Bessel function of the second kind of order $\nu$ \citep{edwards1977unsteady}.

Substituting the expression for $C_M$ and the assumed form of the kinematics into~\eqref{eq:prob5} gives the pitching motion,
\begin{align}
    \theta_{0h} &= \frac{4R\sigma^2 + 2a_0}{-\frac{16}{3}R\sigma^2 + 4K -2a_0 - a_1}h_0^*,
    \label{eq:theta_h} \\
    \theta_{0w} &= \frac{a_w}{-\frac{16}{3}R\sigma_w^2 + 4K -2a_{0w} - a_{1w}}V_w^*. \label{eq:theta_w} 
\end{align}
This gives the pitching motion for the fully coupled fluid-structure interaction problem. The pitch component at reduced frequency $\sigma$ is due to the heaving motion and associated fluid forces, with the wavy flow making no contribution. Conversely, the pitch component at reduced frequency $\sigma_w$ is due to the wavy flow and associated fluid forces, with the heaving motion making no contribution. When $V_w^* = 0$, the kinematics are identical to those in \citet{moore2014analytical}. 

In the special case when the heaving motion has the same frequency as the encounter frequency of the wavy flow ($\sigma = \sigma_w$), the single-frequency pitch amplitude is given by
\begin{equation}
    \theta_0 = \frac{(4R\sigma^2 + 2a_0)h_0^* + a_w V_w^*}{-\frac{16}{3}R\sigma^2 + 4K -2a_0 - a_1},
    \label{eq:theta_t} 
\end{equation}
which is equal to the sum of $\theta_{0h}$ and $\theta_{0w}$ from above. In this special case, the heaving motion, wavy flow, and associated fluid forces all contribute to the single component of the pitch. Although the pitching motion has only one component, we will find it useful to decompose that component into contributions from the heaving motion and from the wavy flow.

\subsection{Measures of propulsive performance}
\label{sec:perf}

With the kinematics solved for, we may calculate various quantities of interest useful for propulsion. The lift coefficient is given by
\begin{equation}
    \label{eq:perf1}
    C_L = \frac{2L}{\rho_f U_\infty^2 cs} = \int_{-1}^1 \Delta p \mathrm{d}x,
\end{equation}
where $L$ is the dimensional lift. The time-averaged power coefficient---giving the average power required to maintain the motion of the foil---is given by
\begin{equation}
    \label{eq:perf2}
    C_P = \frac{2\langle P \rangle}{\rho_f U_\infty^3 cs} = -\int_{-1}^1 \langle \Real(\Delta p) \Real(Y_t) \rangle \mathrm{d} x,
\end{equation}
where $P$ is the dimensional power and $\langle \cdot \rangle$ denotes time-averaging. The time-averaged rate of energy imparted to the fluid is given by
\begin{equation}
    \label{eq:perf3}
    C_E = \frac{2\langle E \rangle}{\rho_f U_\infty^3 cs} = -\int_{-1}^1 \langle \Real(\Delta p) \Real(Y_t + Y_x) \rangle \mathrm{d} x - \langle C_S \rangle
\end{equation}
where $E$ is the dimensional rate of energy imparted to the fluid and $C_S$ is the leading-edge suction coefficient. $E$ is given by the rate of work done by the pressure over the foil's surface \citep{wu1971hydromechanics}. The time-averaged thrust coefficient is then
\begin{equation}
    \label{eq:perf4}
    C_T = \frac{2\langle T \rangle}{\rho_f U_\infty^2 cs} = \int_{-1}^1 \langle \Real(\Delta p) \Real(Y_x) \rangle \mathrm{d} x + \langle C_S \rangle = C_P - C_E,
\end{equation}
where $T$ is the dimensional thrust. Finally, for the propulsive efficiency we use the customary Froude efficiency, which is given by
\begin{equation}
    \label{eq:perf5}
    \eta = \frac{\langle T \rangle U_\infty}{\langle P \rangle} = \frac{C_T}{C_P}.
\end{equation}
Formulae for the above measures of propulsive performance are provided in Appendix~\ref{sec:form}. 
Note that~\eqref{eq:perf4} reflects a fundamental energy balance, 
\begin{equation}
    \label{eq:energy_balance}
    P = TU_\infty + E,
\end{equation}
which holds instantaneously as well as when time-averaged \citep{wu1972extraction}. This balance states that the power input is equal to the sum of the rate of work done by the thrust and the rate of energy imparted to the fluid. The same energy balance holds when there is no wavy component to the incoming flow, in which case $\langle E \rangle \ge 0$ \citep{wu1961swimming}. When the wavy component of the flow is present, however, it is possible for $\langle E \rangle$ to be negative, in which case the foil extracts energy from the wavy flow. 

Although it is possible to have $\langle E \rangle < 0$, this does not necessarily imply that the foil extracts energy from the flow \emph{on net}, which instead requires that $\langle P \rangle < 0$. For all the results shown in this work, the mean thrust is positive, leading to three possible energetic regimes: (1) $0 \le C_E < C_P$ (which gives $0 < \eta \le 1$); (2) $C_E < 0 < C_P$ (which gives $\eta > 1$); and (3) $C_E < C_P < 0$ (which gives $\eta < 0$). In the first regime, the energy supplied to the foil is used to produce thrust and is imparted to the fluid. In the second regime, energy is instead extracted from the wavy flow, but a net input of power is still required to maintain the motion of the foil. The Froude efficiency is greater than unity in this regime because its denominator does not account for the energy available in the flow. In the third regime, the energy extracted from the wavy flow is so great that the motion of the foil extracts net power from the flow ($\langle P \rangle < 0$). As a result, the Froude efficiency is negative in this regime.

\subsection{Parameter space explored}
\label{sec:note}

Since swimmers are thin and neutrally buoyant, they have a low mass ratio $R$. Accordingly, we fix $R = 0.01$ for all the results shown in this work. Fliers have a mass ratio of order unity or higher, so the results presented here may not apply. 

Due to the linearity of the problem, doubling the heave amplitude and wavy velocity amplitude will double the pitch amplitudes and quadruple the mean power, energy imparted to the fluid, and thrust, while leaving the efficiency unchanged. Quantities of interest are therefore reported per unit amplitude. 

The frequency, stiffness ratio, and wavenumber are varied across wide ranges in what follows, even into the range of values unlikely to be encountered in the physical world. The purpose of extending ourselves into very large or very small values of the parameters is to aid in uncovering the key physical principles governing flapping propulsion and heterogeneous flow. Indeed, as we hope the reader will find, asymptotic parameter values aid us in uncovering physical mechanisms whose relevance extends beyond the asymptotic regime.

\section{Dynamics of motion}
\label{subsec:freq_dom}

Making the dependence of the moment coefficient $C_M$ on the heaving and pitching motions explicit in the angular momentum balance in~\eqref{eq:prob5} provides some insight into the dynamics. For heaving motion without waves, the angular momentum balance yields
\begin{equation}
  \label{eq:kin1_1}
  \left( \frac{16}{3}R + \frac{9\upi}{8} \right) \ddot \theta + \frac{3\upi}{2} (C(\textrm{i} \sigma) + 1)\dot \theta + (4K + \upi C(\textrm{i} \sigma)) \theta = (-4R - \upi) \ddot{h}^* - \upi C(\textrm{i} \sigma)\dot h^*,
\end{equation}
and for waves without heaving motion, the angular momentum balance yields
\begin{equation}
  \label{eq:kin2_2}
  \left( \frac{16}{3}R + \frac{9\upi}{8} \right) \ddot \theta + \frac{3\upi}{2} (C(\textrm{i} \sigma_w) + 1)\dot \theta + (4K + \upi C(\textrm{i} \sigma_w)) \theta = a_w V_w^* \textrm{e}^{\textrm{i}\sigma_w t}.
\end{equation}
%These expressions are only well-defined for purely sinusoidal motions due to the presence of Theodorsen's function. 
In both cases, the angular momentum balance takes the form of a standard second-order system with forcing $f(t)$, $\ddot \theta + 2\zeta \omega_0 \dot \theta + \omega_0^2 \theta = f$. Comparing terms, we find that the natural frequency $\omega_0$ and damping ratio $\zeta$ of our system are
\begin{align}
  \label{eq:kin4}
  \omega_0 &= \sqrt{   \frac{ 96K + 24 \upi C}{  128R + 27 \upi     }  }, \\
  \zeta &= \frac{  18 \upi ( C  + 1)   }  {\sqrt{ (96K + 24 \upi C  )( 128R  + 27 \upi} )}  .
\end{align}
Above, the value of $C$ depends on the frequency of the forcing on the right-hand side of the angular momentum balance. The spring and circulatory lift both contribute restoring torques that lead to the oscillator-like behavior. Additionally, both the inertia of the solid and the added mass of the fluid contribute to the total inertia appearing in the denominator of the natural frequency. In contrast, the damping arises solely from the fluid. It may be surprising that the presence of the fluid leads to damping since there is no energy dissipation in the system; however, there is energy transfer between the foil and the fluid which manifests as an effective damping. 

\begin{figure}
\begin{center}
  \includegraphics[width=1\linewidth]{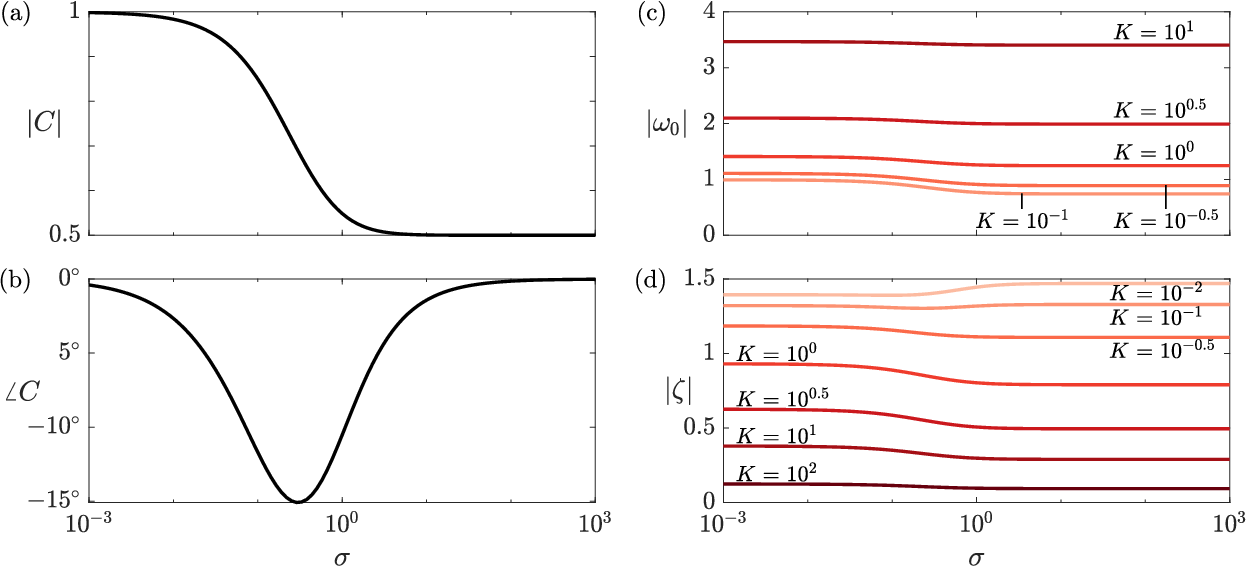}
\end{center}
  \caption{(a) Magnitude of Theodorsen's function, (b) phase of Theodorsen's function, (c) natural frequency, and (d) damping coefficient as a function of the reduced frequency, for $R = 0.01$. }
  \label{fig:theoFreqDamp}
\end{figure}

Due to the presence of Theodorsen's function, the natural frequency and damping ratio are functions of the actuation frequency and are complex numbers. Their magnitudes are shown in figure~\ref{fig:theoFreqDamp}, along with Theodorsen's function. The natural frequency decreases as the actuation frequency increases. However, since the magnitude of Theodorsen's function varies between 1 at low frequencies and $\frac{1}{2}$ at large frequencies, the natural frequency varies little with actuation frequency, especially for large values of the stiffness ratio $K$. How the damping ratio varies with actuation frequency depends on the stiffness ratio, but the variation is small. The natural frequency increases with the stiffness ratio and decreases with the mass ratio $R$, matching our intuition for a mass-spring-damper system. The damping ratio decreases upon increasing the stiffness ratio or mass ratio. The reason is that as $K$ or $R$ increases, the characteristic fluid force becomes comparatively small, thereby making the damping ratio small since damping arises solely from the fluid. 

Taking the Laplace transform of the angular momentum balance equations yields transfer functions describing how the pitching motion depends on the heaving motion and the waves:
\begin{align}
     \frac {\theta_{0h}}{h_0^*}       &= \frac{ (-\upi - 4R) s^2   -  \upi  C(s)s } {\left( \frac{16}{3}R +  \frac{9 \upi}{8} \right) (s^2 + 2\zeta \omega_0 s + \omega_0^2)},    \\
       \frac{\theta_{0w}}{V_w^*}   &=   \frac{ a_w}  {\left( \frac{16}{3}R +  \frac{9 \upi}{8} \right) (s^2 + 2\zeta \omega_0 s + \omega_0^2)}  .
 \end{align}
%Due to the presence of Theodorsen's function, these transfer functions are only well-defined when the Laplace variable $s = \textrm{i}\sigma$ (or $\textrm{i}\sigma_w$). 
The poles of the transfer functions are
\begin{equation}
  \label{eq:kin5}
  s = -\zeta \omega_0 \pm \textrm{i} \omega_0 \sqrt{1 - \zeta^2},
\end{equation}
while the zeros are
\begin{align}
  s &= 0, -\frac{\upi C}{\upi + 4R}, \label{eq:kin6a}\\
  s &= \textrm{i}k^* \left(1 - \frac{W_1 - \textrm{i} W_2}{2J_1 - \textrm{i}J_2} \right), \label{eq:kin6b}
\end{align}
for the heaving motion and waves, respectively. Because the expressions above are frequency-dependent, there is ambiguity in the locations of the poles and zeros. 

For small $R$, the high-frequency limits of the poles (for which $C = \frac{1}{2}$) accurately capture the associated corner frequency (where the transfer function transitions between its low-frequency and high-frequency behavior) and resonant peak at $\omega_0$, so we use the approximation
\begin{equation}
  \label{eq:kin8}
  \omega_0 = \sqrt{\frac{96K + 12\upi}{128R + 27\upi}}.
\end{equation}
Across the natural frequency, the magnitude of the frequency response decreases by a factor of $\sigma^2$ and the phase decreases by $180^\circ$. When the system is overdamped with damping ratio $\zeta > 1$, the transfer function has two poles centered around $-\zeta \omega_0$, with the distance between them increasing with $\zeta$, and there is no resonant peak. In our system, $\zeta$ is bounded from above by $\frac{3}{2}$, so the distance between the poles will be small when our system is overdamped, and the frequency response will be very similar to that of a double-pole at $-\omega_0$. When plotting frequency responses, we will use this simplification and plot $\omega_0$, even in overdamped cases. 

The heave transfer function's second zero depends on Theodorsen's function. Since the imaginary component of Theodorsen's function is always negative and does not acquire large values, the effect that this zero has on the frequency response is quite similar to that of a zero on the negative real axis; that is, the magnitude increases by a factor of $\sigma$ and the phase increases by $90^\circ$ across the corner frequency. Due to the presence of Theodorsen's function, the magnitude and phase both decrease by a small amount before increasing across the corner frequency. For a real-valued zero, the corner frequency is equal to the absolute value of the zero, where the low- and high-frequency asymptotes of the response induced by the zero intersect. Here, we approximate the corner frequency by the magnitude of the high-frequency limit of the zero,
\begin{equation}
 \label{eq:kin9}
  s = -\frac{\upi}{2(\upi + 4R)}.
\end{equation}

 \begin{figure}
 \begin{center}
  \includegraphics[width=\linewidth]{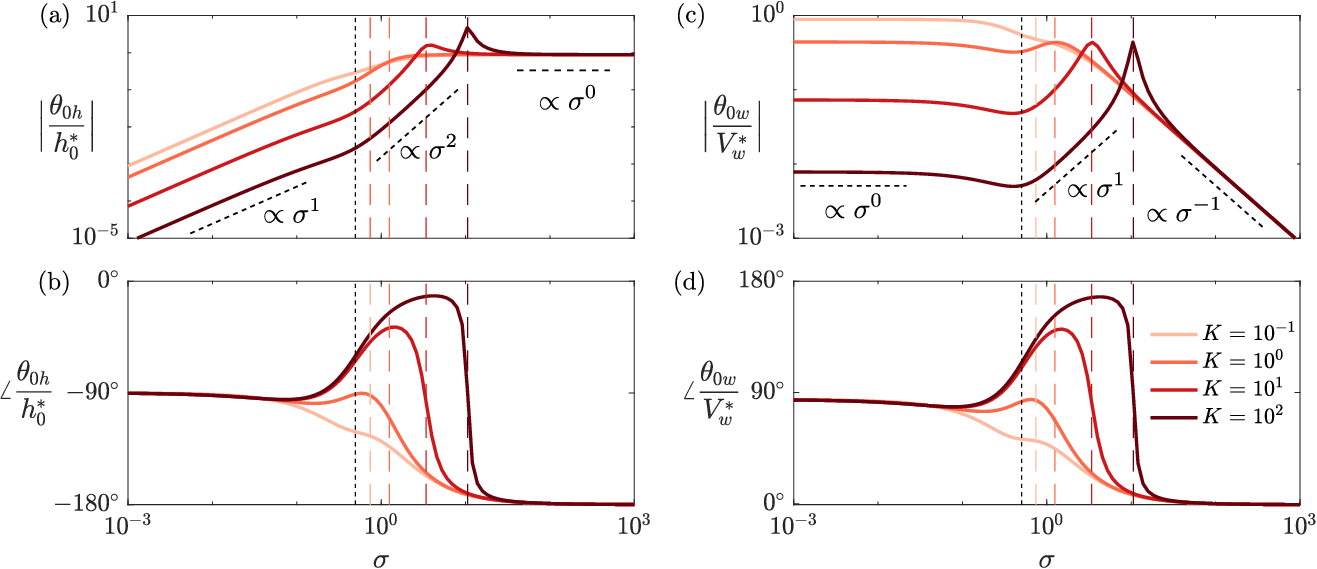}
 \end{center}
  \caption{(a) Magnitude and (b) phase of the response to heaving, for $R = 0.01$. (c) Magnitude and (d) phase of the response to waves, for $R = 0.01$ and $k^* = 0.1$. The zero's location is marked by a vertical short-dashed black line, and the natural frequencies' locations with vertical long-dashed colored lines. }
  \label{fig:bode_heave_wave}
 \end{figure}

The frequency response to the heaving motion is shown in figures~\ref{fig:bode_heave_wave}(a--b). The approximate zero and natural frequency are drawn as dashed vertical lines. For large enough $K$, the damping ratio is small and the frequency response is typical of an underdamped oscillator. At low frequencies, due to the zero at 0, the magnitude of the transfer function is approximately $\frac{\upi}{4K + \upi}\sigma$, and its phase is $-90^\circ$. This phase difference arises since the moment imparted by the fluid is proportional to $-\dot h^*$ at low frequencies, which originates from the quasi-steady circulatory lift. As the frequency crosses the zero and natural frequency, the magnitude and phase change as described previously. There is a resonant peak near the effective natural frequency whose height increases with $K$. The asymptotic behaviors are sketched for large $K$, where the two corner frequencies are well separated. Lower values of $K$ subdue the resonant peak and make the transition between asymptotic behaviors more gradual due to an attendant increase in the damping ratio, with there being no resonant peak for low enough $K$. Additionally, lower values of $K$ push the natural frequency closer toward the zero, so that the regions of transition between asymptotic behaviors overlap significantly. As a result, the magnitude transitions from being $\propto \sigma$ to a constant across the natural frequency, and the phase changes from $-90^\circ$ to $-180^\circ$. Around $\sigma \approx 0.2$, there is a slight deviation in the magnitude and phase from the behavior that was just described due to Theodorsen's function, whose magnitude and phase vary around this frequency. For high frequencies, the response becomes independent of $K$ since inertial forces dominate; the magnitude of the transfer function is approximately $\frac{96R + 24\upi}{128R + 27\upi}$, and its phase is $-180^\circ$. 

For small values of the wavenumber $k^*$, the velocity disturbance induced by the wavy flow is nearly constant across the chord, as it would be for a heaving motion. Consequently, the moment that the wavy flow imparts onto the foil is nearly the same as the moment imparted by the fluid onto a heaving foil (if the vertical component of the wavy velocity is equal to the heave velocity); see Appendix~\ref{sec:special} for details. A difference between these two scenarios is that the heaving motion imparts an inertial moment that scales as $R$, while the wavy flow does not. For the small mass ratios relevant to swimming, this inertial moment is negligible and the frequency response to the waves is nearly the same as that due to the heave \emph{velocity} (although the phase is lower by $90^{\circ}$ due to our convention of multiplying the vertical component of the wavy velocity by i, as described in Section~\ref{sec:prob}, and because the wavy velocity enters the no-penetration boundary condition with a sign opposite to the foil's velocity). This is shown for $k^* = 0.1$ in figures~\ref{fig:bode_heave_wave}(c--d). 

\begin{figure}
 \begin{center}
  \includegraphics[width=0.5\linewidth]{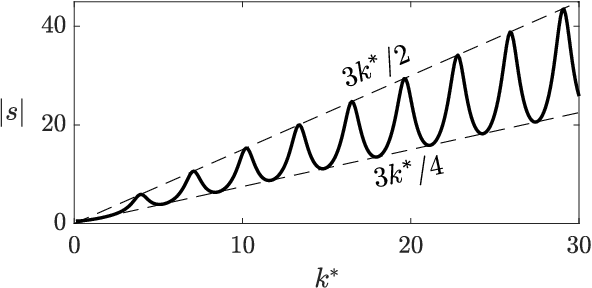}
 \end{center}
  \caption{Magnitude of the approximate zero of the wave transfer function as a function of wavenumber. }
  \label{fig:waveZero}
 \end{figure}

For larger values of the wavenumber, the zero in the frequency response to the waves acquires a significant positive imaginary component while its real component oscillates between negative and positive values. This induces an antiresonant response around frequencies near the magnitude of the zero, where the associated corner frequency is. The effect of the zero is well-approximated by its high-frequency limit, 
\begin{equation}
  s = \textrm{i}k^* \left(1 - \frac{J_1 - \textrm{i} J_0}{4J_1 - 2\textrm{i}J_2} \right).
\end{equation}
The zero has a complicated dependence on $k^*$ due to the oscillatory nature of the Bessel functions, but its magnitude grows with $k^*$ and is very nearly bounded between $\frac{3}{4}k^*$ and $\frac{3}{2}k^*$ (see figure~\ref{fig:waveZero}). The antiresonant response, therefore, occurs when the phase velocity of the waves is between $\frac{3}{4}$ and $\frac{3}{2}$ of the freestream speed (the exact value depending on the wavenumber), for all wavenumbers. 

From a physical standpoint, the range of antiresonant phase velocities can be rationalized by observing that when the phase velocity is equal to the freestream speed ($\sigma = k^*$), the moment (and lift) induced by the non-circulatory flow is exactly zero since the foil moves at a constant velocity through a stationary wavy velocity field. Changing the phase velocity from the freestream speed leads to a non-zero non-circulatory moment that, depending on the wavenumber, may oppose the circulatory moment, thereby diminishing the pitching motion induced by the waves. At the local extrema in figure~\ref{fig:waveZero}, where the non-dimensional phase velocity is $\frac{3}{4}$ or $\frac{3}{2}$, the circulatory and non-circulatory moments cancel and the waves induce no pitching motion. 

\begin{figure}
\begin{center}
  \includegraphics[width=0.8\linewidth]{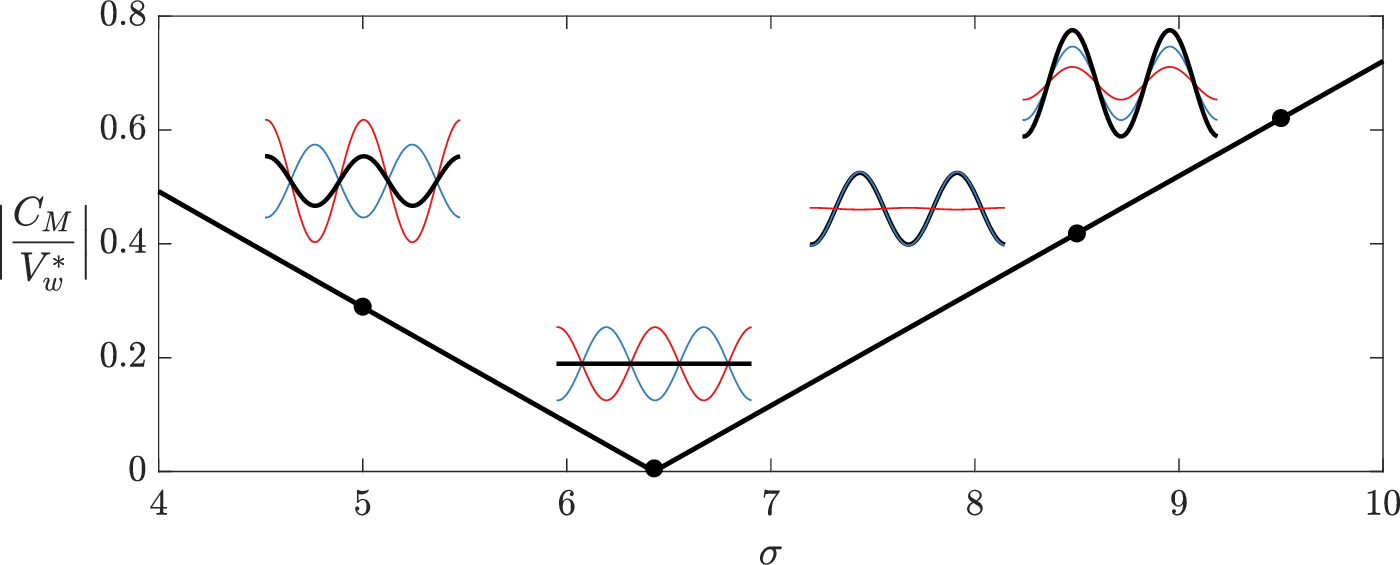}
\end{center}
  \caption{Amplitude of the moment induced by the wavy flow as a function of frequency for $k^* = 8.55$. Insets show the circulatory (blue), non-circulatory (red), and total wavy moments (black) as a function of time at the marked frequencies. }
  \label{fig:momentDecomp}
\end{figure}

Let us consider the case of $k^* = 8.55$, which corresponds to one of the local minima in figure~\ref{fig:waveZero}, in more detail. The moment induced by the wavy flow is shown in figure~\ref{fig:momentDecomp} as a function of frequency, and the insets show the circulatory and non-circulatory components of the moment as functions of time for specific frequencies. At the anti-resonant frequency, the moment is zero. Below the anti-resonant frequency, the circulatory and non-circulatory components of the moment are nearly perfectly out of phase, destructively interfering. As the frequency increases, the magnitude of the non-circulatory moment decreases until it matches that of the circulatory moment, leading to the antiresonant condition. As the frequency increases further, the magnitude of the non-circulatory moment decreases until it becomes zero when the phase velocity is equal to the freestream speed. For higher frequencies, the magnitude of the non-circulatory moment increases, and its phase changes by $180^\circ$. Now, the circulatory and non-circulatory moments constructively interfere.

Perfect cancellation of the circulatory and non-circulatory moments requires that they be perfectly out of phase. This is generally not the case. The phases of the two components of the moment depend strongly on the wavenumber, as does their phase difference. Perfect cancellation is possible only at the wavenumbers corresponding to the local minima and maxima in figure~\ref{fig:waveZero}. However, as long as the two components are not $90^\circ$ out of phase, which is generally true, there will be some degree of cancellation. As explained previously, the maximum degree of cancellation will always be near the frequency where the phase velocity is equal to the freestream speed, across which the non-circulatory moment changes sign. Whether maximum cancellation occurs at a phase velocity above or below the freestream speed depends on the wavenumber, which principally determines the phase difference between the circulatory and non-circulatory moments.

In the high-frequency limit, the numerator of the wave transfer function is
\begin{equation}
  a_w = \upi \left[ -\frac{1}{2} + 2\left( 1 - \frac{\sigma}{k^*} \right) \right] J_1 + \upi \textrm{i} \left[ \frac{1}{2}J_0 - \left( 1 - \frac{\sigma}{k^*} \right)J_2 \right]. 
\end{equation}
Its magnitude is zero in two cases: (1) when the non-dimensional phase velocity $\frac{\sigma}{k^*} = \frac{3}{4}$ and $k^*$ is a root of $J_0 - \frac{1}{2}J_2$, corresponding to the local minima in figure~\ref{fig:waveZero}; and (2) when $k^*$ is a root of $J_1$ and the non-dimensional phase velocity $\frac{\sigma}{k^*} = \frac{3}{2}$, corresponding to the local maxima in figure~\ref{fig:waveZero}. The latter case follows from the recurrence relation for Bessel functions, $J_\nu(k^*) = \frac{k^*}{2\nu} [J_{\nu - 1}(k^*) + J_{\nu + 1}(k^*)]$. These two extremes give the envelope of antiresonant phase velocities. 

 \begin{figure}
 \begin{center}
  \includegraphics[width=\linewidth]{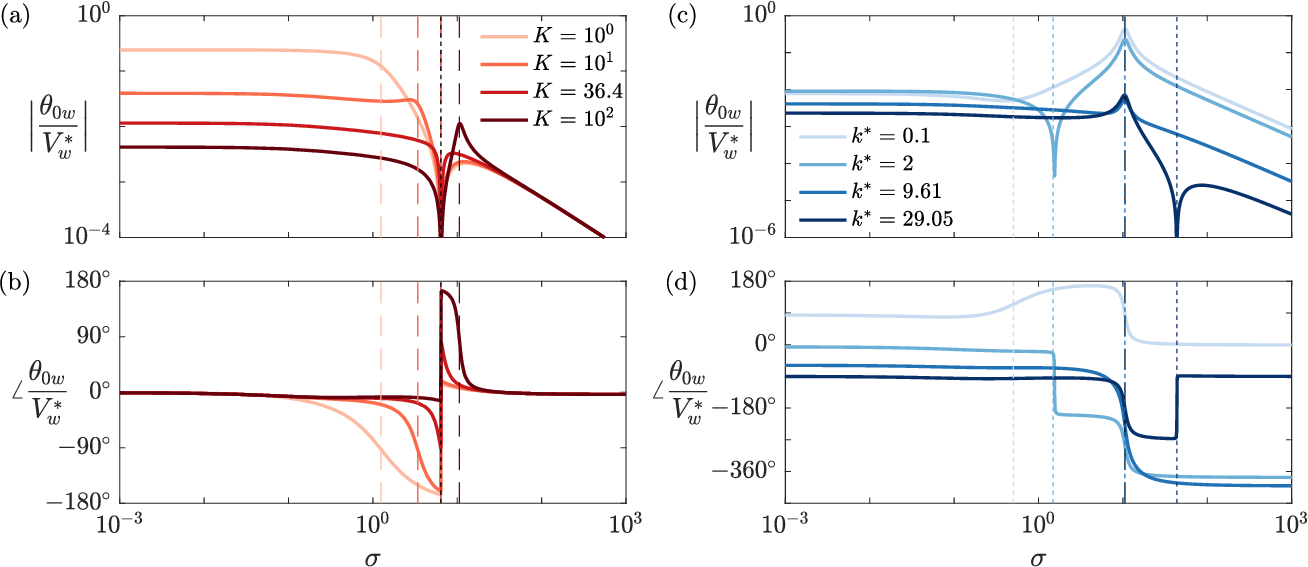}
 \end{center}
  \caption{(a) Magnitude and (b) phase of the response to waves, for $R = 0.01$ and fixed wavenumber $k^* = 8.55$. (c) Magnitude and (d) phase of the response to waves, for $R = 0.01$ and fixed stiffness ratio $K = 100$. In (a) and (b), the natural frequencies' locations are marked by long-dashed colored lines, and the zero's location by a short-dashed black line. In (c) and (d), the zeros' locations are marked by short-dashed colored lines, and the natural frequency's location by a long-dashed black line.  }
  \label{fig:bode_wave}
 \end{figure}

The frequency response to the waves is shown in figure~\ref{fig:bode_wave}. The plots on the left show the frequency response for $k^* = 8.55$, which corresponds to one of the local minima in figure~\ref{fig:waveZero} and a purely imaginary zero. Here, a strong antiresonant response is expected around a non-dimensional phase velocity $\frac{\sigma}{k^*} = \frac{3}{4}$. This is indeed observed, along with an attendant jump of $180^\circ$ in the phase. The antiresonant response is so strong that it can completely overwhelm the resonant response arising from the spring. To illustrate this point, a value of $K = 36.4$ has been chosen for one of the plotted curves since the corresponding natural frequency coincides with the magnitude of the zero. There is not even a hint of resonance in the magnitude plot, although it is still seen in the phase plot. 
 
The plots on the right of figure~\ref{fig:bode_wave} show the frequency response for $K = 100$ and various values of the wavenumber $k^*$. Depending on the wavenumber, the zero can be before or after the natural frequency. When the zero and natural frequency are sufficiently separated, resonant and antiresonant responses appear along one frequency response curve. The zero can also be made to coincide with the natural frequency, such as when $k^* = 9.61$, significantly weakening the resonant response. As a zero is crossed, the magnitude increases by a factor of $\sigma$, as usual, but the change in phase differs drastically compared to when $k^* \ll 1$. Once $k^* \gtrsim 1.3$, the phase of the zero is bounded between approximately $70^\circ$ and $110^\circ$, resulting in changes in the phase of the frequency response between $160^\circ$ and $180^\circ$ across the zero (the phase increases by this amount when the zero is in the left half-plane and decreases when the zero is in the right half-plane). 

At low frequencies, the transfer function is approximately $\upi \frac{2J_1 + \textrm{i}(J_0 - J_2)}{4K + \upi}$, and at high frequencies it is approximately $\frac{24\upi}{\sigma k^*}\frac{2J_1 - \textrm{i} J_2}{128R + 27\upi}$. For high frequencies, the response becomes independent of $K$ since inertial forces dominate. In these limits, in contrast to the response to heave, the phase is non-trivial due to its dependence on the wavenumber. 

\begin{figure}
\begin{center}
  \includegraphics[width=1\linewidth]{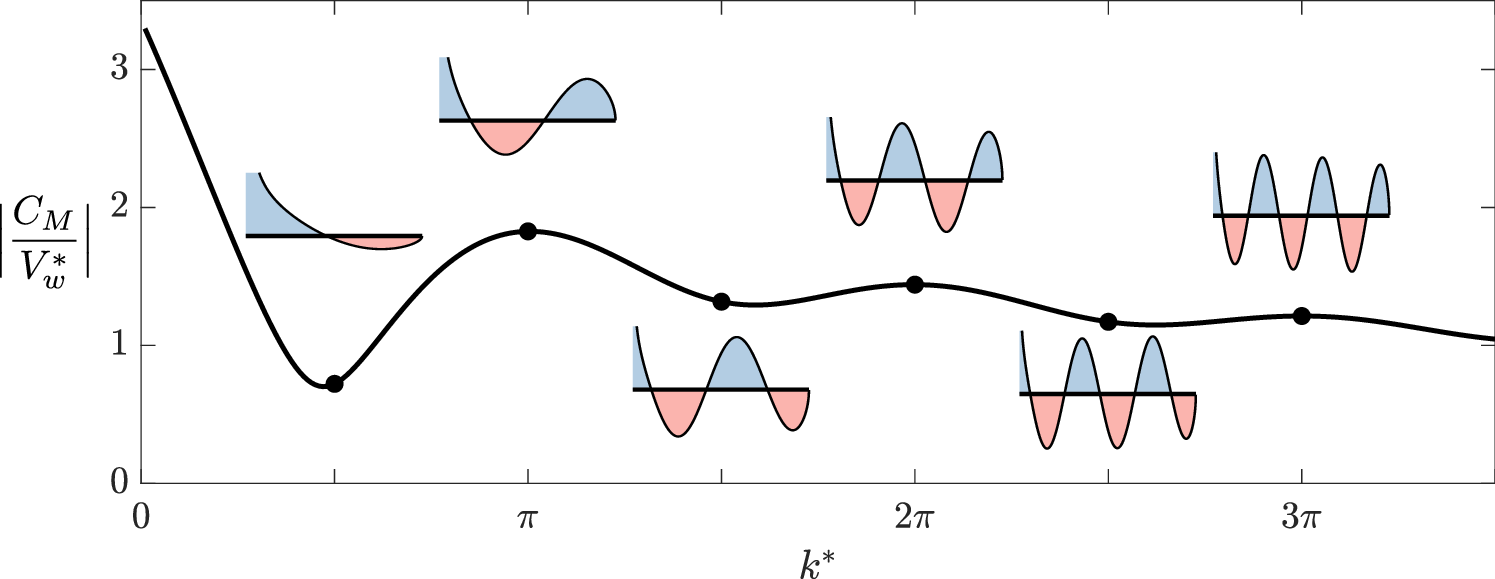}
\end{center}
  \caption{Oscillatory and decaying dependence of coefficient of moment $\vert C_M \vert$ on wavenumber $k^*$ for $K=10$ and $\sigma=1$. Insets show the instantaneous pressure distribution on the foil at the marked points.}
  \label{fig:CM_vs_k}
\end{figure}

Finally, we observe that the magnitude of the frequency response generally weakens as the wavenumber $k^*$ increases. In fact, the magnitude has an oscillatory dependence on $k^*$ on top of a general decay with increasing $k^*$. This is evident in figure~\ref{fig:CM_vs_k}, where the magnitude of the moment imparted by the wavy flow is shown as a function of $k^*$. To explain the dependence on the wavenumber, the distribution of the pressure difference across the foil is shown for various wavenumbers in the same figure. The pressure difference consists of two components: a singular component due to the singularity at the leading edge, and a wave-like component with wavenumber $k^*$. With an increasing number of wavelengths spanning the chord, areas with positive and negative pressure differences increasingly cancel out each other's contribution to the net moment. The singular component of the pressure likewise weakens as the wavenumber increases. This occurs because the singular component of the pressure depends on weighted integrals of the wavy velocity along the chord \citep{wu1972extraction}, and the areas of positive and negative velocity increasingly cancel each other's contribution as the number of wavelengths spanning the chord increases. Since the contributions of the singular and wave-like pressure distributions to the net moment both decrease due to a wave canceling itself, we call this effect destructive self-interference. 

The behavior of the transfer function can be estimated for large $k^*$, for which the Bessel functions are, to leading order, sine or cosine functions of $k^*$ modulated by a $k^{*-1/2}$ envelope. This results in the magnitude of the transfer function being $\propto k^{*-1/2}$, at leading order. This approximation captures the true envelope reasonably well for $k^* \gtrsim 3$. We therefore generally expect a slow $k^{*-1/2}$ decay of the magnitude of the frequency response as $k^*$ increases, though with additional oscillatory dependence on $k^*$. The physical origin of the oscillatory dependence is quite clearly seen in figure~\ref{fig:CM_vs_k}: the degree to which destructive self-interference occurs depends on the fractional number of wavelengths spanning the chord, which is equal to $k^*/\upi$.

\section{Propulsion with unequal heave and wave frequencies}
\label{sec:unequal}

When the heaving motion and wavy flow have different frequencies, the total time-averaged thrust is equal to the sum of the thrust produced by a heaving foil in a uniform flow and a non-heaving foil in a wavy flow: $C_T = C_{T,h} + C_{T,w}$. The same is true for the time-averaged power. Using the angular momentum balance in~\eqref{eq:prob5}, the time-averaged power for a non-heaving foil in a wavy flow can easily be shown to be zero for periodic motion, so that $C_P = C_{P,h}$. The propulsive efficiency is then
\begin{equation}
  \eta = \frac{C_T}{C_P} = \frac{C_{T,h} + C_{T,w}}{C_{P,h}} = \eta_h + \frac{C_{T,w}}{C_{P,h}}.
\end{equation}
Thus, it suffices to separately analyze the propulsive performance of purely heaving and purely wavy motions when their frequencies differ. 

\subsection{Purely heave forcing}
\label{subsec:separate_heave_forcing}

The results for this case are discussed in detail by \cite{moore2014analytical}. We briefly recount some salient features to provide context for later results. 

\begin{figure}
\begin{center}
    \includegraphics[width=\linewidth]{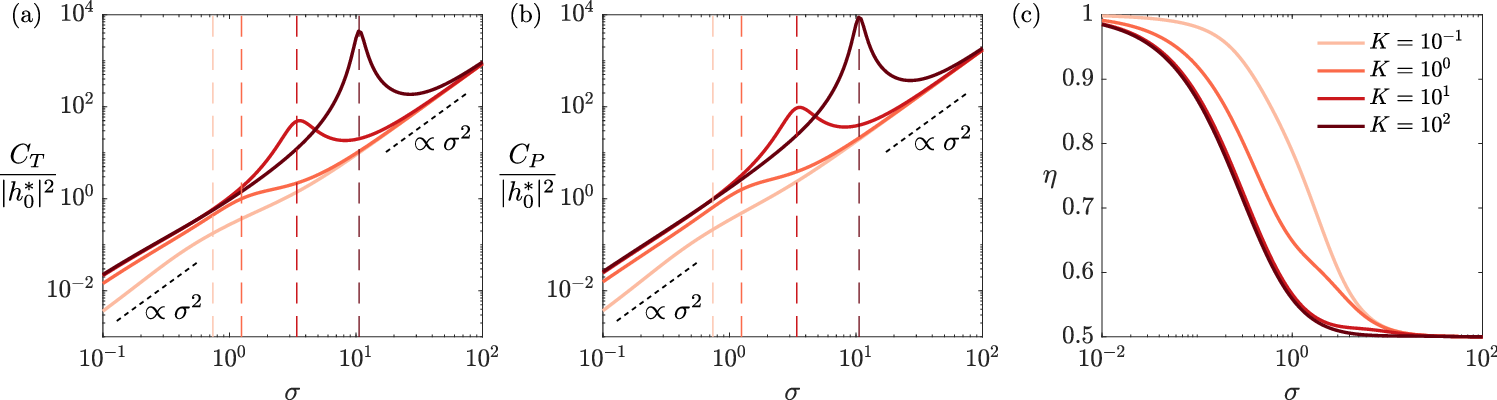}
\end{center}
    \caption{(a) Thrust coefficient, (b) power coefficient, and (c) efficiency of a heaving foil in a uniform flow, for $R = 0.01$. The natural frequencies' locations are marked by vertical long-dashed colored lines.}
    \label{fig:propulsion_heave}
\end{figure}

The variations of thrust, power, and efficiency with frequency and stiffness ratio are shown in figure~\ref{fig:propulsion_heave}. The thrust and power are both $\propto \sigma^2$ in the low- and high-frequency limits, while the efficiency approaches constant values of 1 and $\frac{1}{2}$ in the respective limits. The propulsive performance is independent of the stiffness ratio $K$ in the high-frequency limit due to the independence of the kinematics from $K$ in this limit, while thrust and power increase in the low-frequency limit as the foil stiffens. 

The thrust and power have local peaks around the natural frequency. These resonant peaks are stronger for stiffer foils due to the attendant decrease in damping, and no obvious resonant behavior is apparent for sufficiently flexible foils, as was the case for the pitching motion. The propulsive efficiency, on the other hand, decreases monotonically with frequency, displaying no peaks near the natural frequency. 

\subsection{Purely wavy forcing}
\label{subsec:separate_wavy_forcing}

Since the time-averaged power is zero for purely wavy forcing, we discuss only the thrust. The energy balance in~\eqref{eq:energy_balance} implies that $C_T = -C_E$ in this case, with $C_E$ the average rate of energy imparted to the fluid. On physical grounds, $C_E$ must be negative for the foil to pitch passively (which can also be formally shown \citep{wu1972extraction}), implying that the mean thrust is positive. 

\begin{figure}
\begin{center}
    \includegraphics[width=0.35\linewidth]{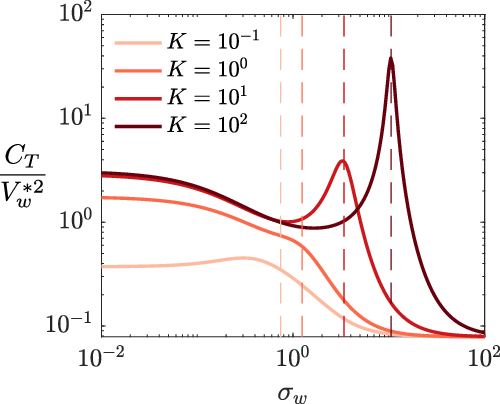}
\end{center}
    \caption{Thrust coefficient of a foil in a wavy flow, for $R = 0.01$ and $k^* = 0.1$. The natural frequencies' locations are marked by vertical long-dashed colored lines.}
    \label{fig:propulsion_wave_ksmall}
\end{figure}

For small values of the wavenumber $k^*$, the average thrust is nearly the same as that produced by a foil heaving in a uniform flow when the heave velocity is equal to the amplitude of the wavy flow; see Appendix~\ref{sec:special} for details. As discussed in Section~\ref{subsec:freq_dom}, small differences arise due to the inertial moment produced by a heaving motion and due to $k^*$ being finite. The thrust coefficient for $k^* = 0.1$ is shown in figure~\ref{fig:propulsion_wave_ksmall}. The curve is essentially the same as that for pure heave in figure~\ref{fig:propulsion_heave}a, except scaled by a factor of $\sigma^{-2}$ since the wavy velocity amplitude is kept constant as $\sigma$ is varied whereas the heave \emph{amplitude} was kept constant as $\sigma$ was varied. 

\begin{figure}
\begin{center}
    \includegraphics[width=0.9\linewidth]{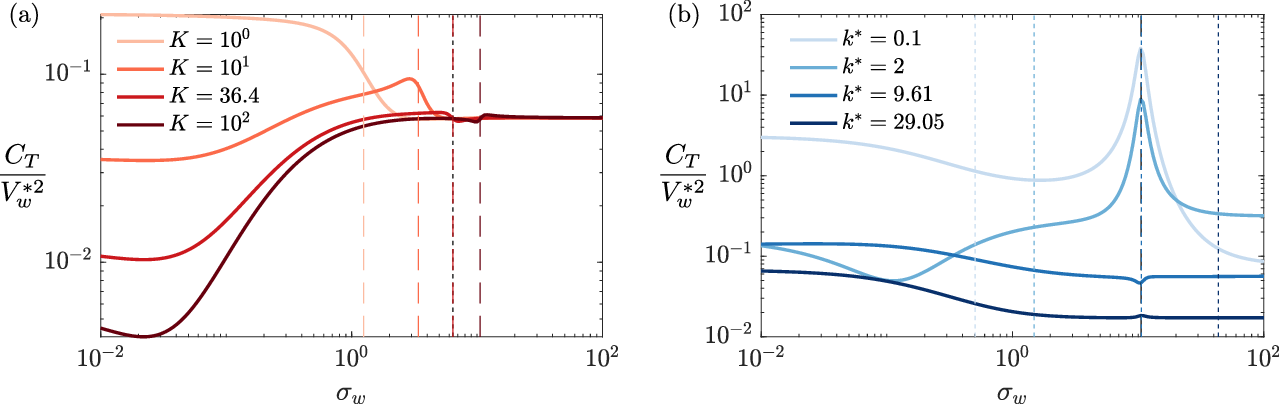}
\end{center}
    \caption{Thrust coefficient of a foil in a wavy flow, for $R = 0.01$ with (a) fixed wavenumber $k^* = 8.55$ and (b) fixed stiffness ratio $K = 100$. In (a), the natural frequencies' locations are marked by long-dashed colored lines, and the zero's location is marked by a short-dashed black line. In (b), the zeros' locations are marked by short-dashed colored lines, and the natural frequency's location is marked by a long-dashed black line.}
    \label{fig:propulsion_wave}
\end{figure}

For larger values of the wavenumber, the possibility of antiresonant behavior arises. The thrust coefficient is shown in figure~\ref{fig:propulsion_wave} for the same parameter values as used for the Bode plots in figure~\ref{fig:bode_wave}. The thrust generally weakens as the wavenumber $k^*$ increases, which we attribute to the destructive self-interference effect discussed in the context of the pitching motion. For large $k^*$, the thrust is oscillatory with a $k^{*-1}$ envelope, to leading order. 

Otherwise, the behavior does not match the intuition that we developed for the pitch response. For a fixed wavenumber of $k^* = 8.55$ (as in figure~\ref{fig:propulsion_wave}a), the pitch response showed a sharp antiresonant trough near the zero, as well as a resonant peak for $K = 100$ near the natural frequency. Neither of these features are present in the thrust. For a fixed stiffness ratio of $K = 100$ (as in figure~\ref{fig:propulsion_wave}b), the pitch response showed antiresonant troughs near the zero and resonant peaks near the natural frequency when the zero and natural frequency were sufficiently separated, and they appeared to cancel each other out when the zero and natural frequency coincided. The thrust, on the other hand, seems to not fully reflect the features of the pitch response. For $k^* = 2$, a resonant peak appears at the natural frequency, but there is no antiresonant trough at the zero. For $k^* = 9.61$, the resonant and antiresonant behavior appear to cancel each other out, as  in the pitch response. For $k^* = 29.05$, neither a resonant peak nor an antiresonant trough are present. 

Decomposing the thrust into its constituents provides clarity. For purely wavy forcing, the mean thrust is the sum of thrust due to pitch ($\propto \vert \theta_{0w} \vert^2$), thrust due to the wavy flow ($\propto V_w^{*2}$), and thrust due to the interaction between the pitching motion and the wavy flow ($\propto \vert \theta_{0w} \vert V_w^*$). The wave thrust is $\upi V_w^{*2} W^2$ (see Appendix~\ref{sec:form}), which is non-negative. Recalling that $C_T = -C_E$, the wave thrust arises from the energy that the wavy flow supplies through the generation of leading-edge suction. Since the wave thrust is independent of the pitching motion, it displays neither resonant nor antiresonant behavior. The pitch thrust is $-\frac{\upi}{4}(F - F^2 - G^2)(9\sigma_w^2 + 4) \vert \theta_{0w}\vert^2$, which is non-positive. Again recalling that $C_T = -C_E$, the non-positivity of the pitch thrust arises from the energy imparted to the fluid as the pitching motion generates a trailing vortex sheet. Being dependent on the pitching motion, the pitch thrust displays resonant and antiresonant behavior. The interaction thrust is more complex, and there are no constraints on the values it can take, other than the total thrust being non-negative. Being dependent on the pitching motion, the interaction thrust displays resonant and antiresonant behavior. 

\begin{figure}
\begin{center}
    \includegraphics[width=0.9\linewidth]{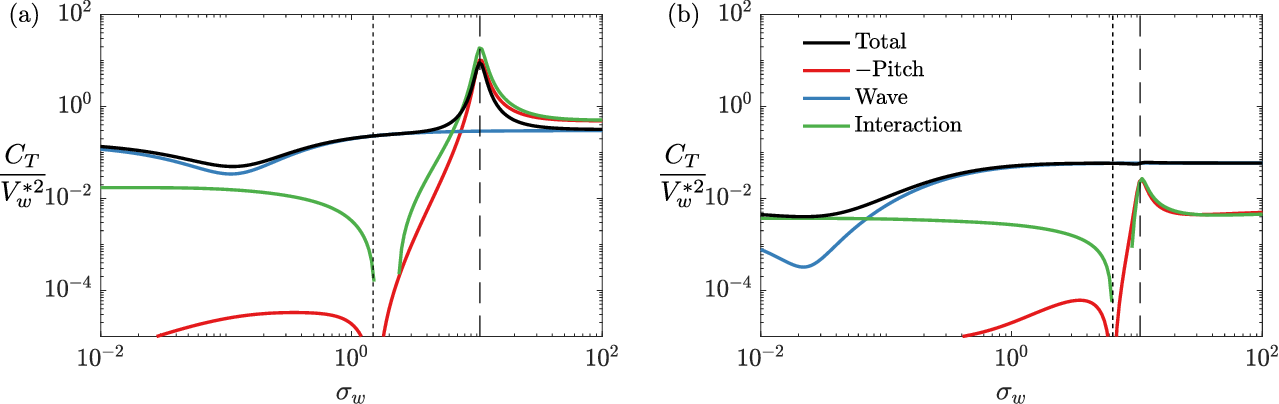}
\end{center}
    \caption{Thrust decomposition for cases (a) $(K, k^*) = (100, 2)$ and (b) $(K, k^*) = (100, 8.55)$ from figure~\ref{fig:propulsion_wave}. }
    \label{fig:propulsion_wave_decomp}
\end{figure}

The decomposition of the thrust into its constituents is shown in figure~\ref{fig:propulsion_wave_decomp} for two cases, with the negative of the pitch thrust plotted. The features of these two cases generally reflect what is seen for the other cases shown in figure~\ref{fig:propulsion_wave}. In the first case ($K = 100$, $k^* = 2$), the pitching motion has a strong antiresonant response at the zero and a strong resonant response at the natural frequency. The antiresonance and resonance of the pitch are apparent in the pitch thrust and interaction thrust. The interaction thrust even becomes negative (though small in absolute value) at frequencies a bit larger than the zero, which is not shown due to the logarithmic ordinate. The wave thrust, on the other hand, is unremarkable across the zero and natural frequency. The wave thrust dominates near the zero, causing the total thrust to show no hints of antiresonance. Near the natural frequency, however, the pitch and interaction thrusts dominate the wave thrust at this wavenumber. Since the interaction thrust is greater in absolute value than the pitch thrust, the total thrust shows a resonant peak that we intuitively expect. 

In the second case ($K = 100$, $k^* = 8.55$), the pitching motion has a strong antiresonant response at the zero and a strong resonant response at the natural frequency. Both are again apparent in the pitch thrust and interaction thrust, while the wave thrust is unremarkable across the zero and natural frequency. As in the previous case, although the pitch thrust and interaction thrust significantly weaken near the zero, the wave thrust does not, causing the total thrust to show no hints of antiresonance. In contrast to the previous case, at this higher wavenumber the pitch and interaction thrusts are weaker than the wave thrust near the natural frequency despite the attendant resonance. Moreover, the pitch and interaction thrusts are nearly equal in absolute value, almost cancelling each other out. As a result, although the sum of the pitch and interaction thrusts has a resonant peak near the natural frequency, it is over an order of magnitude weaker than the wave thrust, leading to practically no hint of resonance in the total thrust. 

We close with a final remark. Since the mean thrust is always positive for purely wavy forcing, the presence of a wavy flow always improves the thrust and propulsive efficiency of a heaving foil when the heaving motion and wavy flow have different frequencies. It is even possible for the efficiency to be greater than unity \citep{wu1972extraction}, though this mostly reflects that the efficiency in~\eqref{eq:perf5} is incorrectly defined since its denominator does not account for the available energy in the wavy flow. Thus, a wavy flow of a different frequency is always beneficial under the assumptions of the linear theory.

\section{Propulsion with equal heave and wave frequencies}
\label{sec:equal_sigma}

When the heaving motion and wavy flow have equal frequencies, the resulting pitching motion is simply a superposition of the pitching motions due to each forcing, so we do not discuss it any further. The thrust and power, however, are quadratic functions of the kinematics, leading to a non-trivial interaction between the two forcings. As a result, we cannot simply consider each forcing in isolation as we could for unequal heave and wave frequencies. 

Having to simultaneously consider both forcings further complicates the matter since two additional non-dimensional parameters become important: the ratio of heave velocity to wave velocity amplitude, and the phase between the heaving motion and wave motion. This expands an already large parameter space. For simplicity, we mostly restrict ourselves to $V_w^* = 1$ and $h_0^* = 1$ while varying the frequency, stiffness ratio, and wavenumber as before. This choice corresponds to the heave and wave velocity having equal strength when $\sigma = 1$, and being in phase at the mid-chord of the foil. For $\sigma < 1$, the wave velocity is stronger, so we expect the propulsive performance to mirror that of a non-heaving foil in a wavy flow (Section~\ref{subsec:separate_wavy_forcing}) for low frequencies. Conversely, we expect the propulsive performance to mirror that of a heaving foil in a uniform flow (Section~\ref{subsec:separate_heave_forcing}) for high frequencies. We also make observations on how the phase difference between the heaving motion and the wave motion, $\phi = \angle h_0^*$, affects the propulsive performance.

\subsection{Long waves}
\label{sec:perflong}

For long waves, the wave velocity enters the boundary-value problem in the same way as the heave velocity, but with opposite sign (see~\eqref{eq:bc}). Approximating the wave velocity as constant across the chord, the two enter the boundary condition as $\textrm{i} \left( \sigma \vert h_0^* \vert \textrm{e}^{\textrm{i} \angle h_0^*} - V_w^* \right) \textrm{e}^{\textrm{i} \sigma t}$. If the heave velocity and wave velocity are in phase and of equal strength ($\sigma \vert h_0^* \vert = V_w^*$), they cancel each other, leading to no pressure difference across the foil when its inertia is negligible. Intuitively, the foil moves with the flow, so we should not expect the flow to exert any force on it. If the heave velocity is stronger than the wave velocity ($\sigma \vert h_0^* \vert > V_w^*$), the wave velocity effectively acts to weaken the heave, so the force felt by the foil will be weaker than that felt by a foil in a uniform flow. If the heave velocity is weaker than the wave velocity ($\sigma \vert h_0^* \vert < V_w^*$), the heave effectively acts to weaken the wave velocity, so the force felt by the foil will be weaker than that felt by a non-heaving foil in a wavy flow. 

\begin{figure}
\begin{center}
    \includegraphics[width=\linewidth]{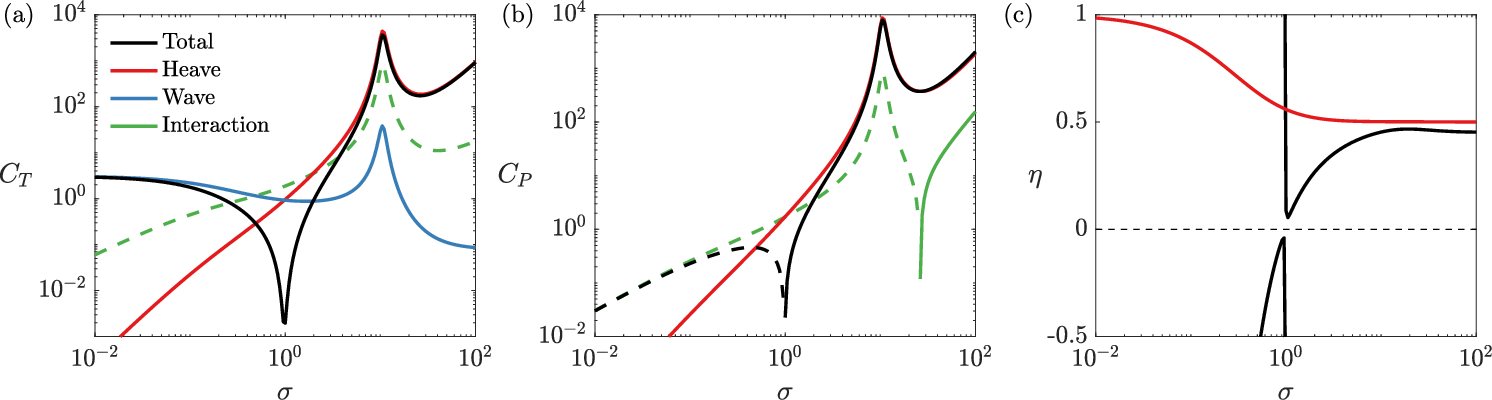}
\end{center}
    \caption{(a) Thrust coefficient, (b) power coefficient, and (c) efficiency of a heaving foil in a wavy flow, for $R = 0.01$, $K = 100$, $k^* = 0.1$, $h_0^* = 1$, and $V_w^* = 1$. Solid portions of curves on logarithmic plots correspond to positive values, while dashed portions correspond to negative values. }
    \label{fig:propulsion_kstarSmall}
\end{figure}

The above arguments are reflected in the average thrust, shown in figure~\ref{fig:propulsion_kstarSmall}a. The thrust is shown only for $K = 100$ since the qualitative features are the same for other stiffness ratios (other than the disappearance of the resonant peak for sufficiently low $K$). Solid portions of curves on logarithmic plots correspond to positive values while dashed portions correspond to negative values. We have decomposed the thrust into thrust due to heave ($\propto \vert h_0^* \vert^2$), thrust due to the wavy flow ($\propto V_w^{*2}$), and thrust due to the interaction between the heaving motion and the wavy flow ($\propto \vert h_0^* \vert V_w^*$). This decomposition differs from that used in Section~\ref{subsec:separate_wavy_forcing}, where we distinguished between thrust arising purely from the wavy flow, purely from the pitching motion, and from the interaction between the wavy flow and pitching motion. Here, these formerly distinct thrust components are grouped together into the wave thrust since they are all proportional to $V_w^{*2}$ due to the linear dependence of $\theta_w$ on $V_w^*$. 

Around $\sigma = 1$, the heave velocity and wave velocity nearly cancel each other at the low wavenumber of $k^* = 0.1$, which manifests as the interaction thrust being nearly equal and opposite to the sum of the heave thrust and wave thrust. For other frequencies, the interaction thrust is always negative, reflecting that the heave velocity and wavy flow always weaken each other when in phase. As a result, in the wave-dominated region (low frequencies), the thrust is a weakened version of the wave thrust, while in the heave-dominated region (high frequencies), the thrust is a weakened version of the heave thrust. 

Since the mean power also depends on the pressure difference across the plate, it is also significantly weakened around $\sigma = 1$ compared to the power for a heaving foil in a uniform flow, as shown in figure~\ref{fig:propulsion_kstarSmall}b. The power can be decomposed in the same way as the thrust. Note that the power due to the wavy flow is zero, as discussed in Section~\ref{subsec:separate_wavy_forcing}, so the total power is a sum of the heave power and interaction power. For low and high frequencies, the behavior of the power differs from that of the thrust. At low frequencies, the mean power is negative; that is, energy is being extracted by the foil. This can be understood by noting that $C_P = - \langle \Real(\dot h^*) \Real(C_L) + \Real(\dot \theta) \Real(C_M) \rangle \approx - \langle \Real(\dot h^*) \Real(C_L) \rangle$, with the error in the approximation being $\textit{O}(R)$. The heave velocity $\dot h^*$ is not changed by the wavy flow, but the lift is. The lift is a sum of the heave lift and wave lift, with the two being $180^\circ$ out of phase and dominated by the wave lift at low frequencies. The sign of the mean power therefore changes compared to that of a foil heaving in a uniform flow once the wave velocity is stronger than the heave velocity ($\sigma < 1$). For sufficiently high frequencies, the interaction power becomes positive. This is due to the wavenumber being finite, and would not occur in the limit $k^* \rightarrow 0$. The power at high frequencies nevertheless approaches that of a heaving plate in a uniform flow since the heave velocity dominates the wave velocity. 

The net effect on the efficiency is that, for $\sigma > 1$, it is reduced below the efficiency in a uniform flow. The efficiency approaches 0.5 in the high-frequency limit, as it does for a heaving foil in a uniform flow (it approaches a lower value near 0.45 in figure~\ref{fig:propulsion_kstarSmall}c due to the wavenumber being finite). For $\sigma < 1$, the efficiency is negative since the power is negative. This negative efficiency reflects that when the heave velocity and wave velocity are in phase and the wave velocity is stronger than the heave velocity, the foil simultaneously produces a net thrust and extracts energy from the flow, which is not possible without waves. 

\begin{figure}
\begin{center}
    \includegraphics[width=\linewidth]{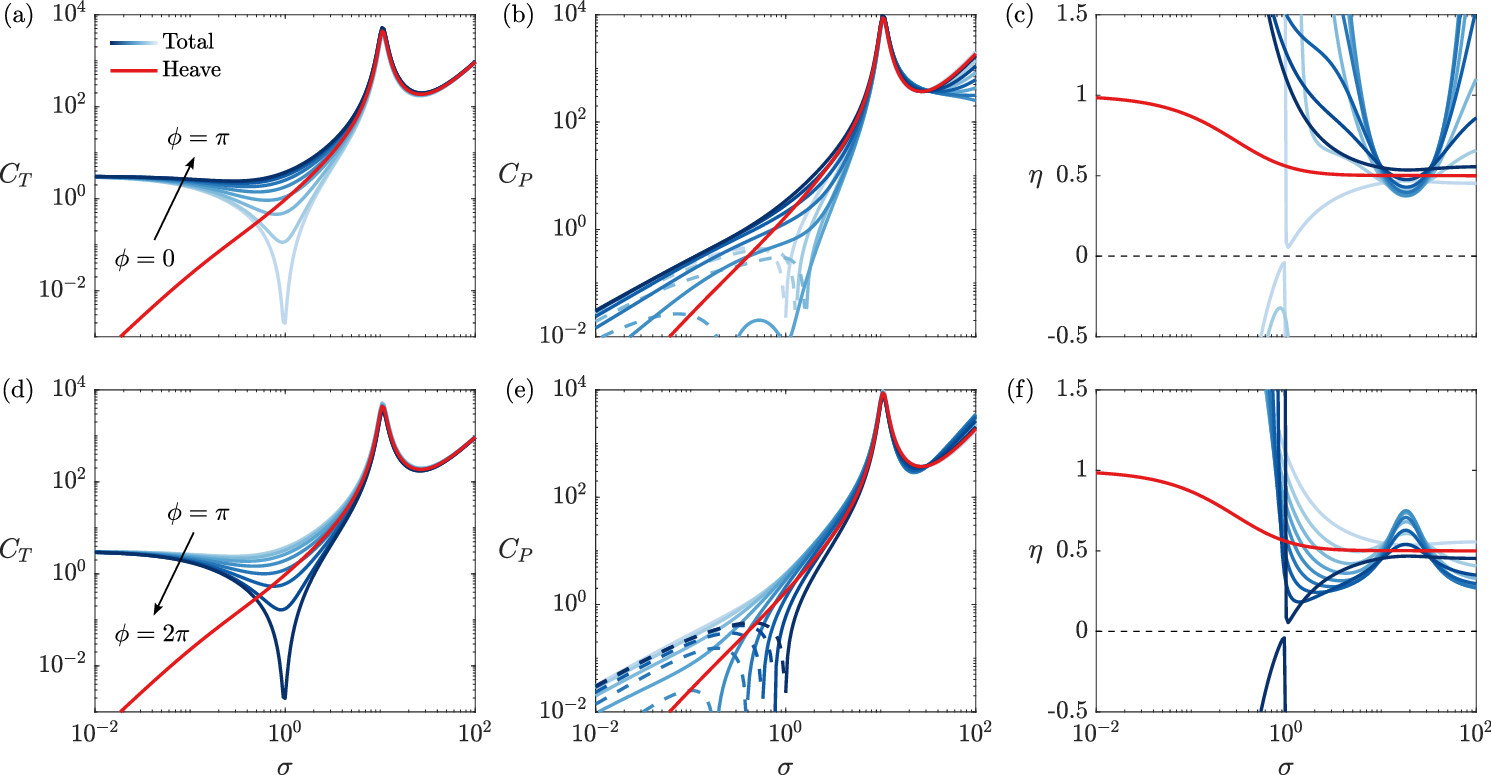}
\end{center}
    \caption{Phase dependence of the (a, d) thrust coefficient, (b, e) power coefficient, and (c, f) efficiency of a heaving foil in a wavy flow, for $R = 0.01$, $K = 100$, $k^* = 0.1$, $\vert h_0^* \vert = 1$, and $V_w^* = 1$. The color darkens as the phase $\phi$ ($= \angle h_0^*$) increases in increments of $\upi/8$. Solid portions of curves on logarithmic plots correspond to positive values, while dashed portions correspond to negative values.} 
    \label{fig:propulsion_phase_kstarSmall}
\end{figure}

For a different value of the phase between the heaving motion and wavy flow, the interaction thrust and power will change while the pure heave and wave components remain unchanged. Notably, the strong cancellation at $\sigma = 1$ only occurs when the heave velocity and wavy velocity are in phase. When the heave velocity and wave velocity are $180^\circ$ out of phase, they work constructively rather than destructively. The resulting interaction thrust has the opposite sign of that in figure~\ref{fig:propulsion_kstarSmall}a, and the total thrust is then greater than the heave thrust and wave thrust. The total thrust is shown for a range of phase differences in figures~\ref{fig:propulsion_phase_kstarSmall}a and d (for clarity, the variation of thrust, power, and efficiency is split into two half-cycles in the top and bottom rows of the figure). Between phases of 0 and $180^\circ$, the thrust varies smoothly and has the same low- and high-frequency behavior. It can be shown that the interaction thrust for a phase of $\phi$ is the negative of that for a phase of $\phi + \upi$.

The interaction power enjoys the same symmetry with respect to phase as the interaction thrust. The dependence of the power on the phase, however, is not as simple, as can be seen in figures~\ref{fig:propulsion_phase_kstarSmall}b and e. As explained previously, at low frequencies the power is dominated by the mean of the product of the wave lift and the heave velocity. In this regime, therefore, the power essentially has a sinusoidal dependence on the phase, and tuning the phase allows us to change the power consumption in a predictable manner with little change to the thrust. A wide range of phases in this regime deliver negative power, that is, net power extraction, while simultaneously producing thrust. For $\sigma \lesssim 1$, all values of the phase deliver efficiency that is either greater than that in pure heave, greater than unity, or even negative (corresponding to simultaneous production of thrust and extraction of power), as shown in figures~\ref{fig:propulsion_phase_kstarSmall}c and f. The behavior appears to be more complex in the large-frequency regime where, despite the heave velocity utterly dominating the wave velocity, the power consumption varies strongly with phase, as does the efficiency. This can be understood by again appealing to the approximation $C_P \approx - \langle \Real(\dot h^*) \Real(C_L) \rangle$. Although the heave lift is much stronger than the wave lift at high frequencies, it is nearly $90^\circ$ out of phase with the heave velocity, making the mean of their product small enough that the interaction power (the mean of the product of the wave lift and heave velocity) can be comparable in magnitude to the heave power. The interaction power depends sinusoidally on the phase, so judicious choice of the phase produces negative interaction power that significantly reduces the total power consumption, leading to increases in the efficiency (and even values of efficiency greater than unity).

\subsection{Finite waves}
\label{sec:fin}

For larger values of the wavenumber $k^*$, at high frequencies (the heave-dominated region) the thrust and power remain close to the thrust and power in a uniform flow when the heave velocity and wave velocity are in phase. At lower frequencies where the wavy flow has greater relative strength, we distinguish between two effects: oscillatory dependence on the wavenumber and a general decay of the effect of the wavy flow as the wavenumber increases. 

\begin{figure}
\begin{center}
    \includegraphics[width=\linewidth]{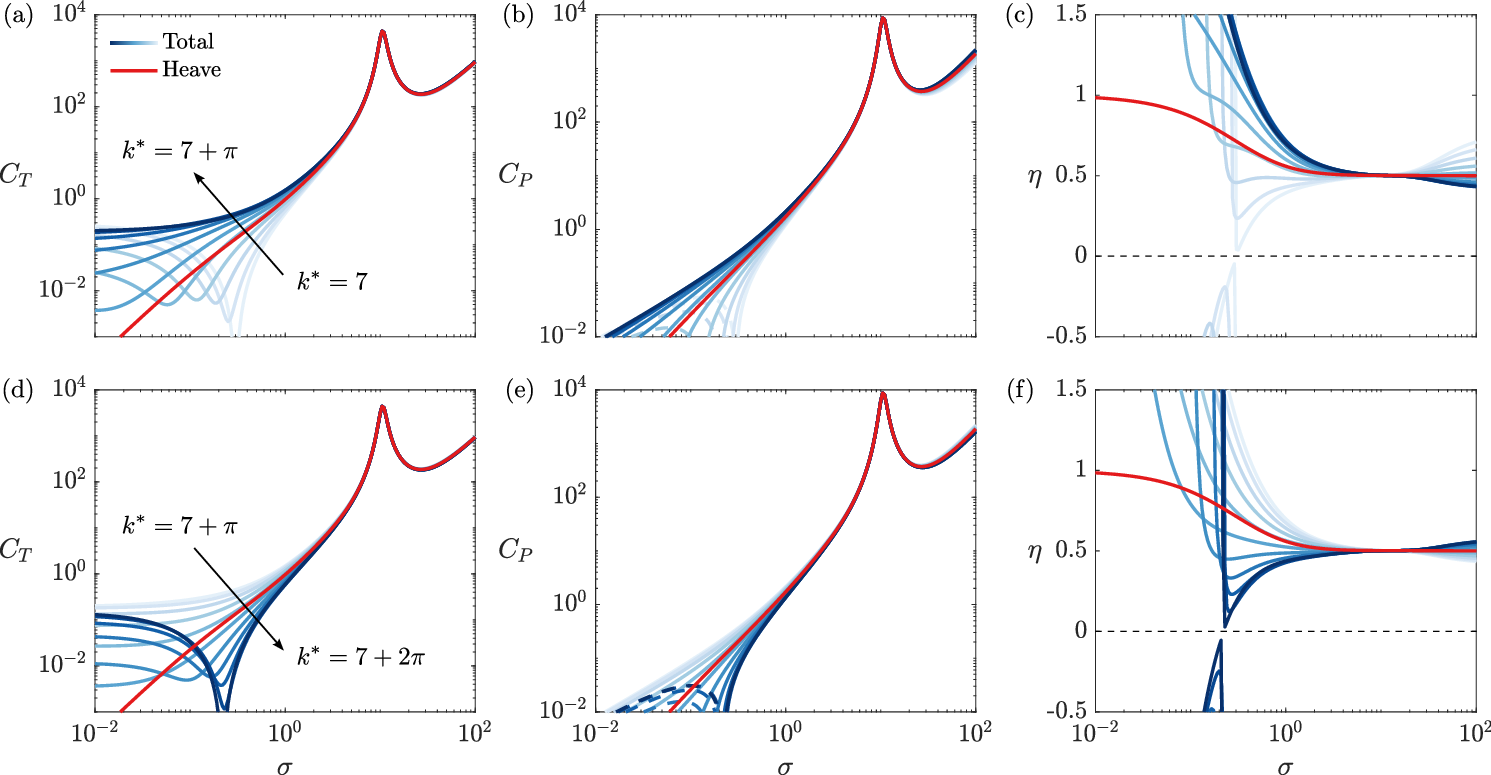}
\end{center}
    \caption{Intra-cycle variation of the (a, d) thrust coefficient, (b, e) power coefficient, and (c, f) efficiency of a heaving foil in a wavy flow, for $R = 0.01$, $K = 100$, $h_0^* = 1$, and $V_w^* = 1$. The color darkens as the wavenumber $k^*$ increases from 7 to $7 + 2\upi$ in increments of $\upi/10$. Solid portions of curves on logarithmic plots correspond to positive values, while dashed portions correspond to negative values. }
    \label{fig:propulsion_kstarUp}
\end{figure}

The oscillatory dependence on $k^*$ is shown in figure~\ref{fig:propulsion_kstarUp}, which shows one characteristic cycle of oscillation for $7 \le k^* \le 7 + 2\upi$. There, the heave velocity and wave velocity are in phase. One cycle has a period of approximately $2\upi$ in $k^*$, based on the large-$k^*$ expansion of the Bessel functions. For clarity, the intra-cycle variation of thrust, power, and efficiency is split into an upward half-cycle (top row of figure~\ref{fig:propulsion_kstarUp}) and a downward half-cycle (bottom row of figure~\ref{fig:propulsion_kstarUp}). For sufficiently low frequencies, the thrust is always greater than that in a uniform flow. In the transition between the heave- and wave-dominated regions, however, the thrust can either be greater or lower than that in a uniform flow, depending on the exact value of the wavenumber. For $k^* = 7$, a strong trough in the thrust is present, as was seen for long waves. As in that case, the interaction thrust is nearly equal and opposite to the sum of the heave thrust and wave thrust. As the wavenumber increases, the interaction thrust weakens in the transition region, eventually becoming positive and leading to an increase in thrust over that in a uniform flow. The maximal benefit is reached half of a cycle later, and further increases of the wavenumber nearly reverse the process, with the thrust showing a strong trough for $k^* = 7 + 2\upi$. This process continues to repeat cyclically as the wavenumber increases. Although this particular range of wavenumbers contains a strong anti-resonance at $(k^*, \sigma) = (8.55, 6.41)$, its effects are not seen for two reasons: the antiresonant frequency falls in the heave-dominated regime, and purely heaving effects are not subject to antiresonance; and even the wavy contribution does not display an antiresonant effect, as was explained by figure~\ref{fig:propulsion_wave_decomp}. 

The intra-cycle variation of the power is similar. For low frequencies, the power expended is lowest for $k^* = 7$ and increases as $k^*$ increases during the first half of the cycle before decreasing during the second half of the cycle. The regions of negative average power change similarly, being strongest during the initial part of the cycle before disappearing and eventually returning. 

The intra-cycle variation of the efficiency follows directly from that of the thrust and power. The beginning of the cycle at $k^* = 7$ produces increases in the efficiency at high frequencies, modest decreases at moderate frequencies, sizable decreases for reduced frequencies of order 1 and smaller, before producing negative efficiency at low frequencies due to net power extraction at low frequencies. The other half of the cycle yields the opposite effect. Notably, the half-cycle centered about $k^* = 7 + \upi$ yields a wide swath of frequencies $\sigma \lesssim 10$ with greater efficiency than in a uniform flow, with sufficiently low frequencies yielding efficiency greater than unity. 

\begin{figure}
\begin{center}
    \includegraphics[width=\linewidth]{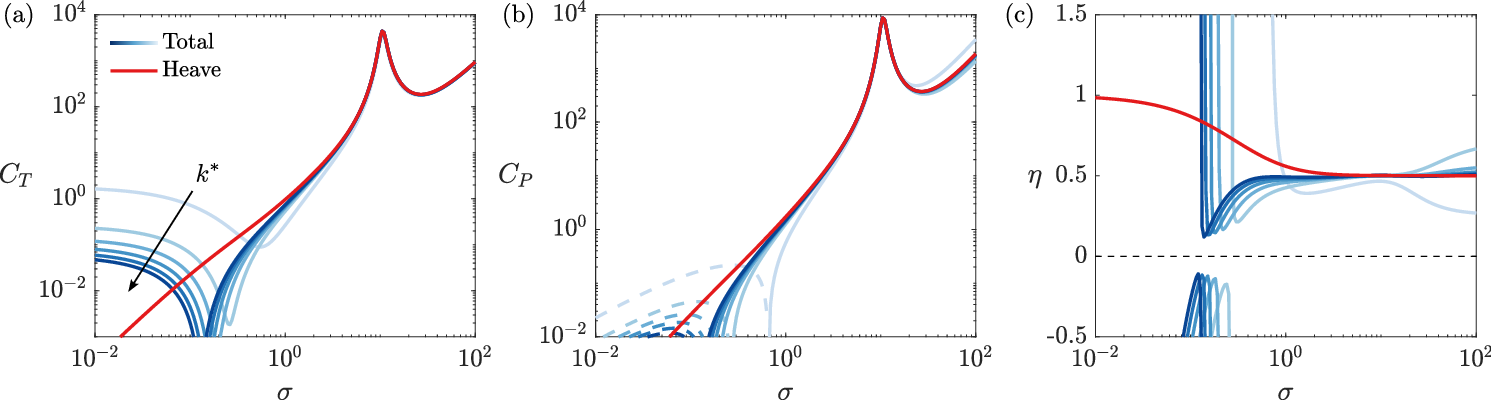}
\end{center}
    \caption{Inter-cycle variation of the (a) thrust coefficient, (b) power coefficient, and (c) efficiency of a heaving foil in a wavy flow, for $R = 0.01$, $K = 100$, $h_0^* = 1$, and $V_w^* = 1$. The color darkens as the wavenumber $k^*$ increases from 1 to $1 + 10\upi$ in increments of $2\upi$. Solid portions of curves on logarithmic plots correspond to positive values, while dashed portions correspond to negative values. }
    \label{fig:propulsion_intercycle}
\end{figure}

With increasing wavenumber, there is a general decay of the effect of the wavy flow. This is most clearly seen by plotting the inter-cycle variation of thrust, power, and efficiency, which is done in figure~\ref{fig:propulsion_intercycle} by sampling the wavenumber at a rate of $2\upi$. The effect of the wavy flow on the thrust and power evidently decreases as the wavenumber increases, with the wave-dominated regime weakening, the transition to the wave-dominated regime being to pushed to lower frequencies, and the propulsive performance decaying closer to the propulsive performance for pure heave elsewhere. Although the inter-cycle variation is shown for only one phase within the cycle, we have verified that the trends just described hold for other wavenumbers as well. Note that the decay occurs at a decreasing marginal rate, reflecting that the pressure distribution induced by the wavy flow destructively interferes with itself at a decreasing marginal rate. 

\begin{figure}
\begin{center}
    \includegraphics[width=\linewidth]{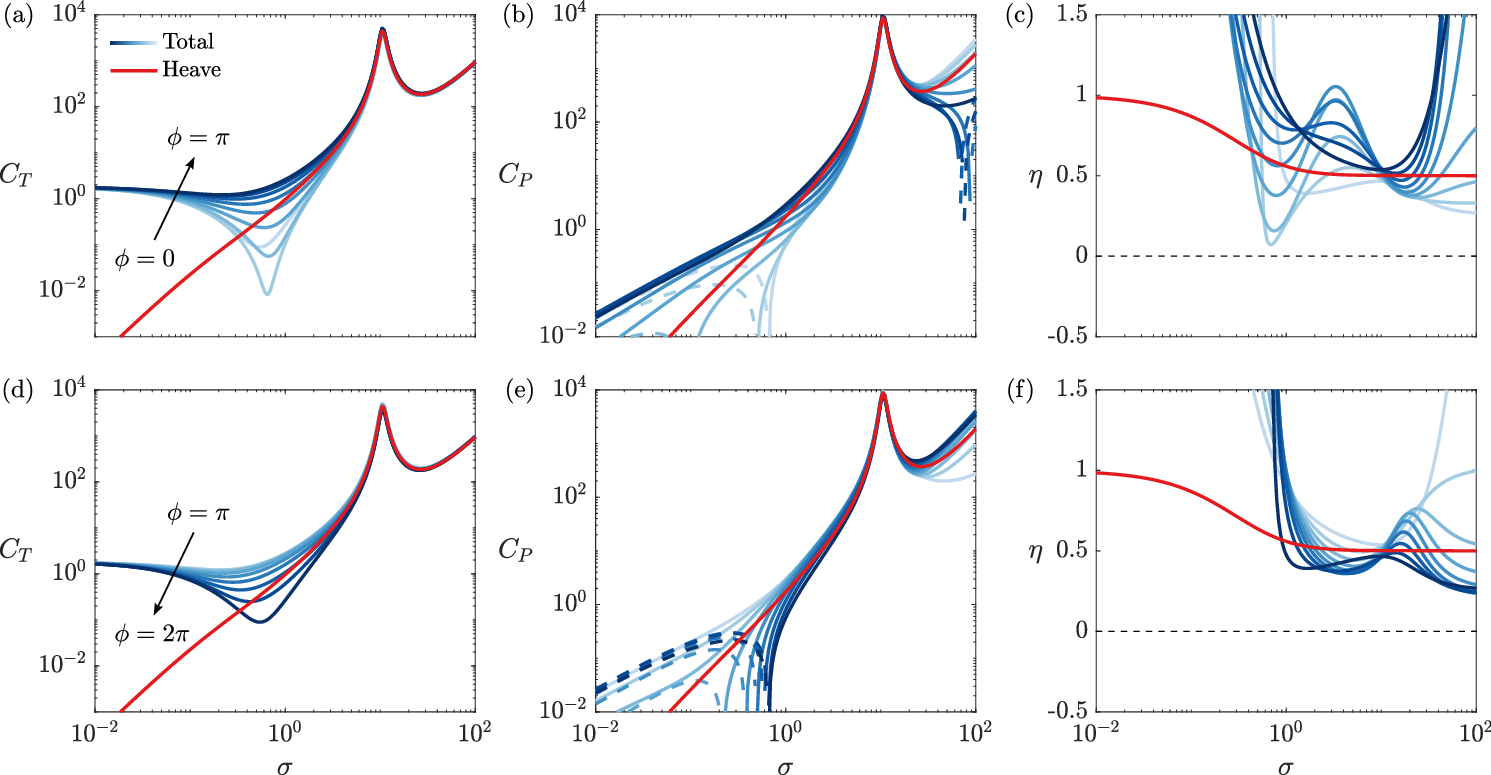}
\end{center}
    \caption{Phase dependence of the (a, d) thrust coefficient, (b, e) power coefficient, and (c, f) efficiency of a heaving foil in a wavy flow, for $R = 0.01$, $K = 100$, $k^* = 1$, $\vert h_0^* \vert = 1$, and $V_w^* = 1$. The color darkens as the phase $\phi$ ($= \angle h_0^*$) increases in increments of $\upi/8$. Solid portions of curves on logarithmic plots correspond to positive values, while dashed portions correspond to negative values.}
    \label{fig:propulsion_phase}
\end{figure}

For finite waves, the variations of thrust, power, and efficiency with the phase difference between the heaving motion and wavy flow are qualitatively the same as for long waves. As shown in figure~\ref{fig:propulsion_phase} for $k^* = 1$, the thrust approaches the heave thrust for high frequencies and the wave thrust for low frequencies. A phase difference of 0 leads to the strong cancellation of the heave velocity and wave velocity near $\sigma = 1$, while for a phase difference of $180^\circ$ they work constructively. The thrust varies smoothly for other phase differences. As the wavenumber increases, the location of the thrust trough generally shifts to lower frequencies due to the general decay of the effect of the wavy flow with increasing wavenumber that was noted earlier. The power and efficiency behave similarly as in the case of long waves, but with one exception. At high frequencies, certain values of the phase difference produce negative net power. For $k^* = O(1)$, the wave lift has greater magnitude than it does for long waves, leading to greater interaction power that enables net power extraction for certain phase differences. For greater values of the wavenumber, however, the wave lift weakens due to the destructive self-interference effect of the pressure waves, and net power extraction is not attained. The other features of how propulsive performance varies with phase persist at higher wavenumbers.

\subsection{Special case of zero pitch}

When the heaving motion and waves have equal frequencies, the pitching motion can be made to have any magnitude and phase relative to the wavy velocity by judicious choice of the heaving motion. We investigate the special case where there is no passive pitching motion, achieved by setting $h_0^* = -a_w V_w^*/(4R\sigma^2 + 2a_0)$.

The transfer function describing how the required heaving motion depends on the waves is
\begin{equation}
  \frac{h_0^*}{V_w^*} = \frac{a_w}{(\upi + 4R)s^2 + \upi C(s)s}.
\end{equation}
This heaving motion is independent of the elastic restoring torque of the spring, but it does depend on the foil's inertia (which is negligible for the small mass ratios relevant to swimming). The poles of the transfer function are
\begin{equation}
  s = 0, -\frac{\upi C}{\upi + 4R}
\end{equation}
and its zero is
\begin{equation}
  s = \textrm{i}k^* \left(1 - \frac{W_1 - \textrm{i} W_2}{2J_1 - \textrm{i}J_2} \right).
\end{equation}
They are respectively identical to the zeros of the pitch transfer functions in~\eqref{eq:kin6a} and~\eqref{eq:kin6b}, and we approximate the corner frequencies as before. 

\begin{figure}
 \begin{center}
  \includegraphics[width=\linewidth]{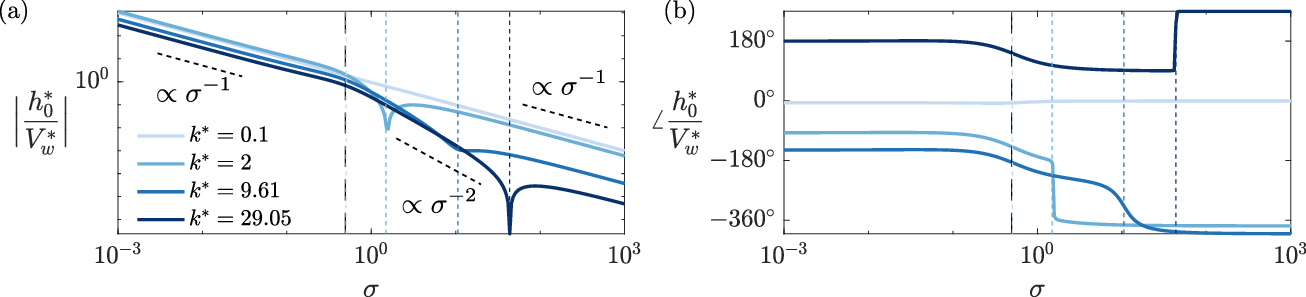}
 \end{center}
  \caption{(a) Magnitude and (b) phase of the heave required for zero pitch, for $R = 0.01$. The pole's location is marked by a long-dashed black line, and the zeros' locations by short-dashed colored lines. }
  \label{fig:kinematicsZeroPitch}
 \end{figure}

The magnitude and phase of the required heaving motion are shown in figure~\ref{fig:kinematicsZeroPitch}. The approximate pole and zero are drawn as dashed vertical lines. As described previously, long waves enjoy a near equivalence to heaving, so negating the effect of the wave simply requires a heaving motion with equal and opposite velocity. Hence the heave amplitude is proportional to $\sigma^{-1}$ (with a constant of proportionality near unity) with a phase near $0^\circ$ for low values of the wavenumber. 

For larger values of the wavenumber, the same basic intuition holds for low and high frequencies. The required heave velocity is proportional to the wave velocity, though with a constant of proportionality that decreases as the wavenumber increases since the pressure distribution induced by the wave destructively interferes with itself. The phase of the moment induced by the wave will also vary as the number of wavelengths spanning the chord changes. 

The behavior differs for non-asymptotic frequencies. Across the pole, the magnitude decreases by a factor of $\sigma$ and the phase decreases by $90^\circ$. This is clear for sufficiently large $k^*$ since the pole and zero are well separated, as can be seen for the case where $k^* = 29.05$ in figure~\ref{fig:kinematicsZeroPitch}. The zero may have a significant positive imaginary component, inducing an antiresonant response near the magnitude of the zero along with an attendant jump in the phase. Since the zero is identical to that of the pitch response to the waves in Section~\ref{subsec:freq_dom}, so is the behavior it induces. The strength of the antiresonance depends on the damping of the zero; for example, the zero has a relatively much smaller real part for $k^* = 29.05$ than for $k^* = 9.61$. The direction of the jump in the phase, the precise low-frequency behavior, and the precise high-frequency behavior have non-trivial dependence on the wavenumber due to the Bessel functions. We do not discuss this dependence in detail since there is little additional physical insight to gain. 

\begin{figure}
\begin{center}
    \includegraphics[width=\linewidth]{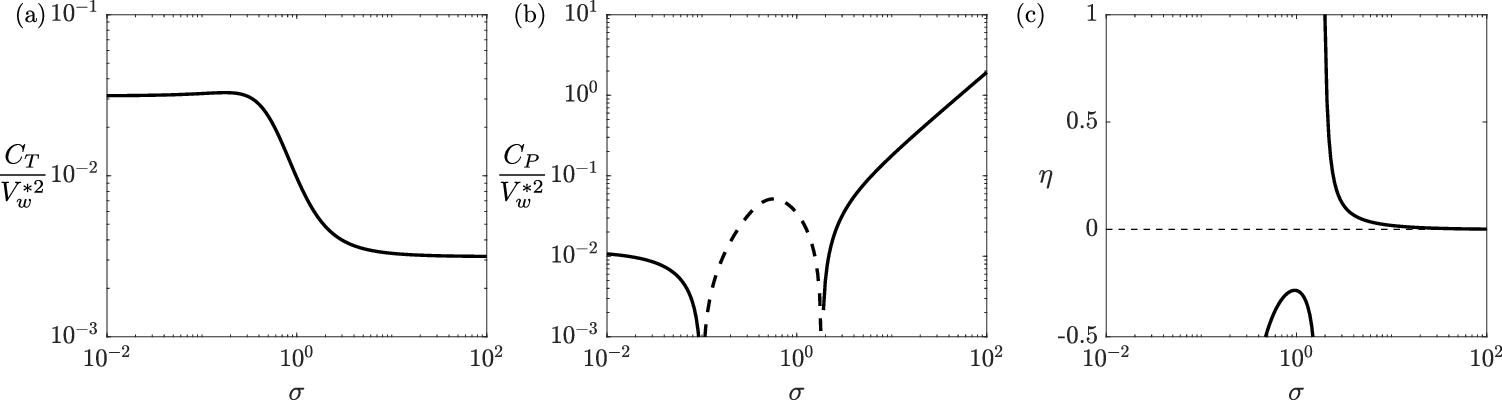}
\end{center}
    \caption{(a) Thrust coefficient, (b) power coefficient, and (c) efficiency of a heaving foil in a wavy flow, for $R = 0.01$, $k^* = 0.1$, $V_w^* = 1$, and heave that produces zero pitch. Solid portions of curves on logarithmic plots correspond to positive values, while dashed portions correspond to negative values. }
    \label{fig:propulsion_kstarSmall_zeroPitch}
\end{figure}

Since the kinematics are independent of the stiffness ratio $K$, so are the measures of propulsive performance. The thrust, power, and efficiency for long waves ($k^* = 0.1$) are shown in figure~\ref{fig:propulsion_kstarSmall_zeroPitch}. The most important features are the low values of the thrust and power. Although not shown for clarity, the heave and wave thrust are at least an order of magnitude greater than the total thrust. The disparity in magnitude widens for greater stiffness ratios. Moreover, the heave and wave thrust have resonant peaks, while the total thrust does not. The interaction thrust nearly cancels the heave and wave thrusts at low wavenumbers. The physical explanation for the greatly diminished thrust is the same as for the steep trough in thrust in figure~\ref{fig:propulsion_kstarSmall}a. Likewise, the interaction power nearly cancels the heave power, both of which are at least an order of magnitude greater than the total power for $\sigma \lesssim 1$, and both of which have resonant peaks despite the total power not reflecting a resonance. The physical explanation is the same as that for the steep trough in power in figure~\ref{fig:propulsion_kstarSmall}b. For higher frequencies, the power increases linearly, but this is due to the wavenumber being finite. Similarly, the variation in efficiency can be attributed to the wavenumber being finite. In the limit $k^* \rightarrow 0$, the thrust and power tend to zero when the inertia of the foil is negligible, with the thrust approaching zero faster, causing the efficiency to also approach zero. 

\begin{figure}
\begin{center}
    \includegraphics[width=\linewidth]{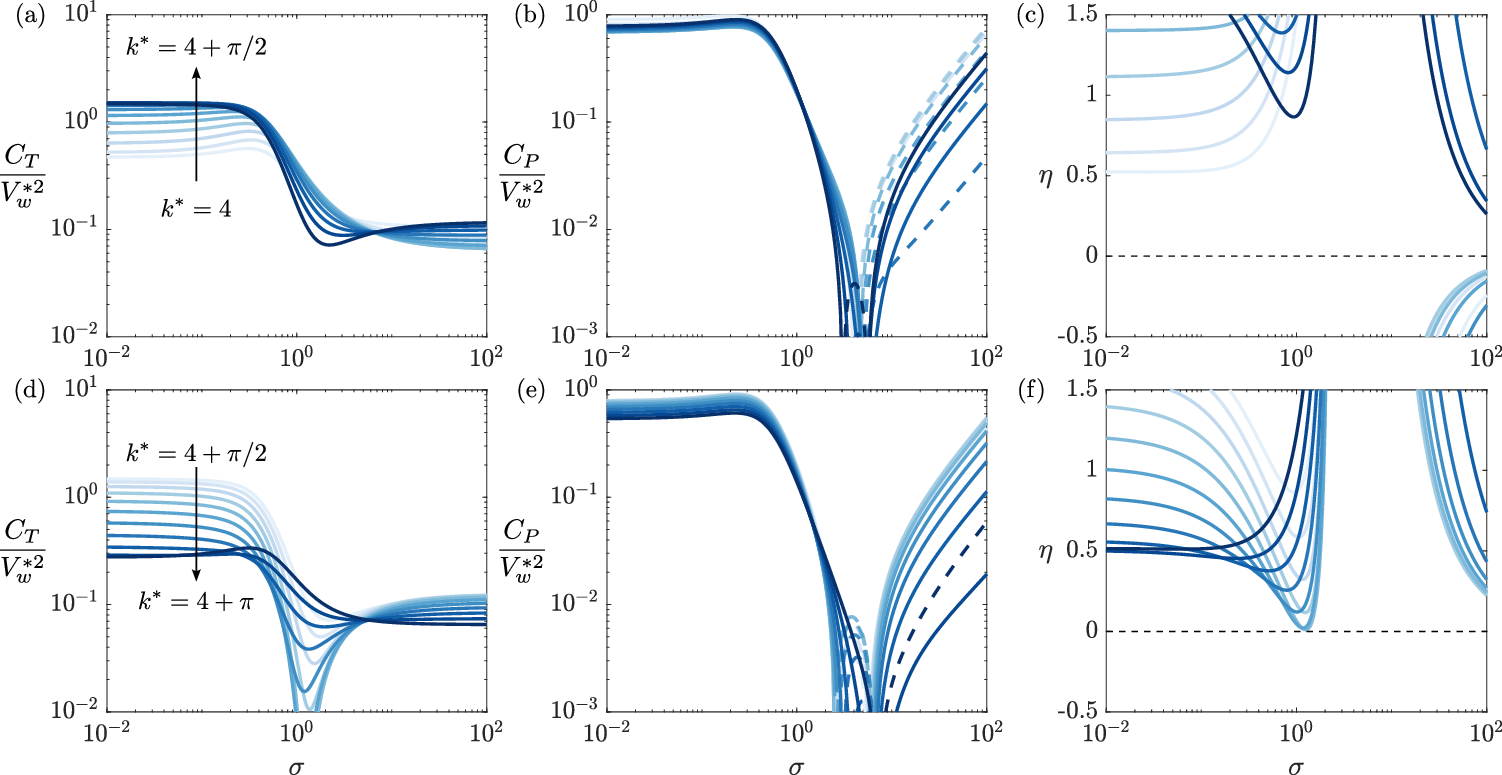}
\end{center}
    \caption{Intra-cycle variation of the (a, d) thrust coefficient, (b, e) power coefficient, and (c, f) efficiency of a heaving foil in a wavy flow, for $R = 0.01$, $V_w^* = 1$, and heave that produces zero pitch. The color darkens as the wavenumber $k^*$ increases from 4 to $4 + \upi$ in increments of $\upi/20$. Solid portions of curves on logarithmic plots correspond to positive values, while dashed portions correspond to negative values. }
    \label{fig:propulsion_kstarUp_zeroPitch}
\end{figure}

For larger values of the wavenumber, the thrust and power have an oscillatory dependence on the wavenumber on top of a general decay with increasing wavenumber. These two effects are qualitatively the same as described in Section~\ref{sec:fin}. The oscillatory dependence on $k^*$ is shown in figure~\ref{fig:propulsion_kstarUp_zeroPitch} for one characteristic cycle of oscillation over $4 \le k^* \le 4 + \upi$. 

In contrast to when the heaving motion was independent of the waves, the cycle now has a period of $\upi$ in $k^*$ instead of $2\upi$. Since the thrust and power are quadratic functions of the kinematics and wavy flow, and the cyclic part of the Bessel functions has a period of $2\upi$, the wave thrust and power always have a period of $\upi$ in $k^*$. When the heaving motion is independent of the wavy flow, the heave thrust and power are independent of $k^*$, and the interaction thrust and power depend on $k^*$ only through the wavy flow, therefore having a period of $2\upi$ in $k^*$. When the heaving motion depends on the wavy flow---as it must in order for the pitching motion to be zero---the Bessel functions appear in the heaving motion, rendering the heave and interaction thrust and power quadratic functions of the Bessel functions, and therefore having period $\upi$ in $k^*$. 

As the wavenumber varies within the cycle, the thrust develops a strong trough. This trough is unrelated to the antiresonance present in the heave transfer function. Although the heave and interaction thrust experience a significant decrease around the zero of the transfer function, the wave thrust does not, so neither does the total thrust, similar to what arose in Section~\ref{subsec:separate_wavy_forcing}. The cyclic trough that appears in the total thrust is instead a result of the interaction thrust becoming large and negative at certain points of the cycle. 

The power, on the other hand, has a persistent region where it weakens significantly and can even become negative. This region is close to the antiresonant zero of the heave transfer function, though it does not exactly coincide with it. It is, however, the antiresonant zero that is responsible for this persistent trough. The heave power has an antiresonant trough at the zero's location, and the interaction power weakens significantly around the zero, though not exactly at the zero since the interaction power does not depend purely on the heaving motion. Since the wave power is zero, the total power weakens significantly near the zero. 

Finally, the efficiency varies significantly within a cycle, from being quite low in a narrow range of frequencies to being quite large and even greater than unity over a large range of frequencies. There are large ranges of frequencies where the efficiency is negative, corresponding to regions where positive thrust is produced while energy is on net extracted from the flow. The main difference from the previously considered heaving motion is that the current motion enables simultaneous net energy extraction and positive thrust even at high frequencies.

\section{Conclusions}
\label{sec:conclusion}

In this work, we studied the dynamics and propulsion of a flexible flapping foil in a wavy flow. We used the so-called torsional flexibility model, in which a rigid foil is attached to a torsional spring at its leading edge \citep{moore2014analytical}. The torsional spring allows the foil to pitch passively in response to moments from the prescribed heaving motion and fluid flow. The fluid flow took the form of a generic traveling wave disturbance superposed on a uniform freestream velocity. Using the linear inviscid flow model of \citet{wu1972extraction}, valid for small-amplitude motions with attached flow, enabled an analytical approach via which several physical insights were drawn. The mass ratio was fixed to a low value representative of swimmers, while the other non-dimensional parameters spanned a large range of values. 

In response to the prescribed heaving motion, the elastic and fluid forces induce a well-known resonant response typical of damped oscillators. The response to the wavy flow is essentially the same when the waves are long since the incoming flow experienced by the foil is nearly spatially uniform, as it is when the foil is heaved. 

Shorter waves, on the other hand, give rise to a number of new physical effects. First, the response has a strong oscillatory dependence on the wavenumber. Second, the response tends to weaken as the wavenumber increases. Together, these two effects reflect an oscillatory dependence on the wavenumber with a decaying envelope. This arises from destructive self-interference of the wave. The wavy flow induces a wave-like pressure distribution along the foil, and areas with positive and negative pressure differences increasingly cancel out each other's contribution to the total response as an increasing number of wavelengths span the length of the foil. In effect, the foil behaves as a low-pass filter. The oscillatory dependence on the wavenumber arises since a fractional number of wavelengths may span the foil. The third new physical effect is the appearance of an antiresonance that leads to sharp minima in the response to the wavy flow. The antiresonant response occurs when the phase velocity of the waves is between $\frac{3}{4}$ and $\frac{3}{2}$ of the freestream speed, which gives rise to non-circulatory forces that weaken the circulatory forces. At certain wavenumbers spaced $\frac{\upi}{2}$ apart, the antiresonant effect is especially pronounced and can even completely overwhelm the resonant response arising from the elastic forces. 

When the wavy flow is of a different frequency than the prescribed heaving motion, it always serves to increase the thrust and efficiency of propulsion relative to a uniform flow. This reflects a conversion of energy available in the incoming flow into useful work done by the thrust force, which has been observed in experiments \citep{beal2006passive}. When the wavy flow and heaving motion are of the same frequency, the net effect on propulsion is nuanced and strongly depends on the wavenumber and phase difference between the wavy flow and heave, both of which determine whether the two forcings weaken or reinforce each other. The relative strengths of the heave and wave velocity amplitude determine which forcing dominates the propulsive response. It is possible to simultaneously produce a positive thrust force while extracting energy on net from the flow. 

The fundamental insights provided in this work extend beyond simple traveling wave flow disturbances to generic spatially and temporally heterogeneous flows since any sufficiently well-behaved disturbance can be expanded into a superposition of traveling waves. The linearity of our model renders the analysis of multi-wave disturbances simple. Although all frequencies and wavenumbers will contribute to the resulting pitch dynamics and propulsive performance, longer waves will tend to have stronger contributions due to the low-pass nature of the foil. In other words, large-scale flow structures predominantly determine the disturbance-induced response. This is beneficial from a practical point of view since if one would like to predict the response to an incoming disturbance, only the large-scale structure of the disturbance need be known to develop an accurate prediction. The level of detail with which a disturbance must be known to develop an accurate prediction is further decreased by the tendency of realistic flow disturbances to have a decaying spectrum.

We conclude with some thoughts about finite Reynolds number effects, which our model, being inviscid, does not account for. A finite Reynolds number primarily manifests as a negative offset to the thrust coefficient \citep{senturk2019reynolds}. Such an offset drag would lead to frequencies across which the net thrust transitions from negative to positive values. In uniform flows, this tends to create a local maximum in the efficiency near the transition \citep{floryan2017scaling, fernandez2017note, floryan2018efficient}, as well as near natural frequencies for flexible foils \citep{floryan2018clarifying}. In wavy flows, the effect on the efficiency will be more complex. Letting $\eta_0 = \frac{C_T}{C_P}$ denote the efficiency in the absence of the offset drag $C_D$, accounting for the offset drag would change the efficiency to $\eta = \eta_0 - \frac{C_D}{C_P}$. Because the power coefficient has a non-trivial dependence on the wave parameters, so will the change in the efficiency. However, we may gain some insight by appealing to the energy balance, which can be written as $C_T = C_P - C_E$ in the inviscid case. Offsetting the thrust coefficient by a fixed value is equivalent to increasing $C_E$---the rate of energy imparted to the fluid---by a fixed amount. This equivalence is reasonable from a physical standpoint since the presence of viscosity would lead to energy dissipation. We expect that increased dissipation would increase the effective damping in the angular momentum balance, weakening the resonant peaks in the kinematics, but not leading to qualitative differences in the kinematics. The physical effects of anti-resonance and destructive interference should persist in a viscous flow, just as the effect of resonance does. Detailed insights on the effects of a finite Reynolds number must await further work, computational or experimental.

\begin{bmhead}[Declaration of interests.]
The authors report no conflict of interest.
\end{bmhead}

% \appendix
\begin{appen}

\section{Formulae for measures of propulsive performance}
\label{sec:form}

Here, we provide formulae for the measures of propulsive performance introduced in Section~\ref{sec:perf}. 

\subsection{Case $\sigma \neq \sigma_w$}
\label{sec:form1}

The lift coefficient is
\begin{equation}
    \label{eq:form1_1}
    C_L = e^{i\sigma t}(\beta_0 b_0 + \beta_1 b_1) + e^{i\sigma_w t}(\beta_{0w}b_{0w} + \beta_{1w}b_{1w} + V_w^* b_w),
\end{equation}
where
\begin{align}
    \label{eq:form1_2}
    b_0 &= -\upi i \sigma C(\textrm{i} \sigma) + \frac{\upi \sigma^2}{2}, \\
    b_1 &= -2\upi C(\textrm{i} \sigma) - \upi i \sigma C(\textrm{i} \sigma) - \upi i \sigma, \\
    b_{0w} &= -\upi i \sigma_w C(\textrm{i} \sigma_w) + \frac{\upi \sigma_w^2}{2}, \\
    b_{1w} &= -2\upi C(\textrm{i} \sigma_w) - \upi i \sigma_w C(\textrm{i} \sigma_w) - \upi i \sigma_w \\
    b_w &= -2\upi [W_1(\textrm{i} \sigma_w) - i W_2(\textrm{i} \sigma_w)] + 2\upi \left( 1 - \frac{\sigma_w}{k^*} \right) J_1(k^*).
\end{align}
The time-averaged power coefficient is
\begin{equation}
\begin{aligned}
    \label{eq:form1_3}
    C_P =& \frac{\upi\sigma}{4} \left[ P_{11}(\sigma)|\beta_0|^2 + P_{33}(\sigma)|\beta_1|^2 + 2P_{13}(\sigma)\Re(\beta_0 \overline{\beta_1}) + 2P_{14}(\sigma)\Im(-\beta_0 \overline{\beta_1}) \right] \\
    &+ \frac{\upi\sigma_w}{4} \left[ P_{11}(\sigma_w)|\beta_{0w}|^2 + P_{33}(\sigma_w)|\beta_{1w}|^2 + 2P_{13}(\sigma_w)\Re(\beta_{0w} \overline{\beta_{1w}}) + 2P_{14}(\sigma_w)\Im(-\beta_{0w} \overline{\beta_{1w}}) \right] \\
    &+ \frac{\upi\sigma_w}{2} V_w\left[ P_1'(\sigma_w)\Re(\beta_{0w}) + P_2'(\sigma_w)\Im(\beta_{0w}) + P_3'(\sigma_w)\Re(\beta_{1w}) + P_4'(\sigma_w)\Im(\beta_{1w}) \right],
\end{aligned}
\end{equation}
where an overbar denotes complex conjugation and 
\begin{align}
    \label{eq:form1_4}
    P_{11}(\sigma) &= \sigma F(\textrm{i} \sigma), \\
    P_{13}(\sigma) &= G(\textrm{i} \sigma) + \frac{\sigma}{2}, \\
    P_{14}(\sigma) &= F(\textrm{i} \sigma) - \sigma G(\textrm{i} \sigma), \\
    P_{33}(\sigma) &= \sigma[1 - F(\textrm{i} \sigma)] - 2G(\textrm{i} \sigma) \\
    P_1'(\sigma_w) &= -W_2(\textrm{i} \sigma_w), \\
    P_2'(\sigma_w) &= \left(1 - \frac{\sigma_w}{k^*}\right)J_1(k^*) - W_1(\textrm{i} \sigma_w), \\
    P_3'(\sigma_w) &= W_2(\textrm{i} \sigma_w) + \left(1 - \frac{\sigma_w}{k^*}\right)J_2(k^*), \\
    P_4'(\sigma_w) &= W_1(\textrm{i} \sigma_w).
\end{align}
The time-averaged rate of energy imparted to the fluid is
\begin{equation}
\begin{aligned}
    \label{eq:form1_5}
    C_E =& \frac{\upi}{4}B(\sigma)\left\{ Q_{11}(\sigma)[|\beta_0|^2 + 2\Re(\beta_0 \overline{\beta_1})] + Q_{33}(\sigma)|\beta_1|^2 + 2Q_{14}(\sigma)\Im(-\beta_0 \overline{\beta_1}) \right\} \\
    &+ \frac{\upi}{4}B(\sigma_w)\left\{ Q_{11}(\sigma_w)[|\beta_{0w}|^2 + 2\Re(\beta_{0w} \overline{\beta_{1w}})] + Q_{33}(\sigma_w)|\beta_{1w}|^2 + 2Q_{14}(\sigma_w)\Im(-\beta_{0w} \overline{\beta_{1w}}) \right\} \\
    &+ \frac{\upi}{2} V_w^*\left[ Q_1'(\sigma_w)\Re(\beta_{0w}) + Q_2'(\sigma_w)\Im(\beta_{0w}) + Q_3'(\sigma_w)\Re(\beta_{1w}) + Q_4'(\sigma_w)\Im(\beta_{1w}) \right] \\
    &- \upi V_w^{*2} W(\sigma_w)^2,
\end{aligned}
\end{equation}
where
\begin{align}
    \label{eq:form1_6}
    B(\sigma) &= F(\textrm{i} \sigma) - [F(\textrm{i} \sigma)^2 + G(\textrm{i} \sigma)^2], \\
    Q_{11}(\sigma) &= \sigma^2, \\
    Q_{14}(\sigma) &= 2\sigma, \\
    Q_{33}(\sigma) &= \sigma^2 + 4, \\
    Q_1'(\sigma_w) &= \sigma_w\left[W_2(\textrm{i} \sigma_w) - 2B(\sigma_w)J_0(k^*)\right], \\
    Q_2'(\sigma_w) &= \sigma_w\left[2B(\sigma_w)J_1(k^*) + P_2'(\sigma_w)\right], \\
    Q_3'(\sigma_w) &= \sigma_w P_3'(\sigma_w) - 2P_2'(\sigma_w) - 2[\sigma_w J_0(k^*) + 2J_1(k^*)]B(\sigma_w), \\
    Q_4'(\sigma_w) &= 2W_2(\textrm{i} \sigma_w) - \sigma_w W_1(\textrm{i} \sigma_w) - 2[2J_0(k^*) - \sigma_w J_1(k^*)]B(\sigma_w), \\
    W(\sigma_w)^2 &= W_1(\textrm{i} \sigma_w)^2 + W_2(\textrm{i} \sigma_w)^2.
\end{align}
The time-averaged thrust coefficient is then
\begin{equation}
    \label{eq:form1_7}
    C_T = C_P - C_E.
\end{equation}

\subsection{Case $\sigma = \sigma_w$}
\label{sec:form2}

The lift coefficient is unchanged. The time-averaged power coefficient is
\begin{equation}
\begin{aligned}
    \label{eq:form2_2}
    C_P =& \frac{\upi\sigma}{4} \Bigr[ P_{11}(\sigma)|\beta_0 + \beta_{0w}|^2 + P_{33}(\sigma)|\beta_1 + \beta_{1w}|^2 + 2P_{13}(\sigma)\Re((\beta_0 + \beta_{0w})\overline{(\beta_1 + \beta_{1w})}) \\
    &+ 2P_{14}(\sigma)\Im(-(\beta_0 + \beta_{0w}) \overline{(\beta_1 + \beta_{1w})}) \Bigr] \\
    &+ \frac{\upi\sigma}{2} V_w^*\left[ P_1'(\sigma)\Re(\beta_0 + \beta_{0w}) + P_2'(\sigma)\Im(\beta_0 + \beta_{0w}) + P_3'(\sigma)\Re(\beta_1 + \beta_{1w}) + P_4'(\sigma)\Im(\beta_1 + \beta_{1w}) \right].
\end{aligned}
\end{equation}

The time-averaged rate of energy imparted to the fluid is
\begin{equation}
\begin{aligned}
    \label{eq:form2_3}
    C_E =& \frac{\upi}{4}B(\sigma)\Bigr\{ Q_{11}(\sigma)[|\beta_0 + \beta_{0w}|^2 + 2\Re((\beta_0 + \beta_{0w}) \overline{(\beta_1 + \beta_{1w})})] + Q_{33}(\sigma)|\beta_1 + \beta_{1w}|^2 \\
    &+ 2Q_{14}(\sigma)\Im(-(\beta_0 + \beta_{0w}) \overline{(\beta_1 + \beta_{1w})}) \Bigr\} \\
    &+ \frac{\upi}{2} V_w^*\left[ Q_1'(\sigma)\Re(\beta_0 + \beta_{0w}) + Q_2'(\sigma)\Im(\beta_0 + \beta_{0w}) + Q_3'(\sigma)\Re(\beta_1 + \beta_{1w}) + Q_4'(\sigma)\Im(\beta_1 + \beta_{1w}) \right] \\
    &- \upi V_w^{*2} W(\sigma)^2.
\end{aligned}
\end{equation}

The time-averaged thrust coefficient is then
\begin{equation}
    \label{eq:form2_4}
    C_T = C_P - C_E.
\end{equation}

\section{Long and short waves}
\label{sec:special}

\subsection{Long-wave limit}
\label{sec:long}

As $k^* \rightarrow 0$, the wavelength of the wavy velocity becomes much longer than the chord, so the wavy velocity field will be nearly constant along the plate. On physical grounds, we therefore expect the wavy velocity to have the same effect as a heaving motion. Expanding the wavy velocity in powers of $k^*$ yields
\begin{equation}
    \label{eq:long1}
    v_w(x,t) = i V_w^* e^{i\sigma_w t} + V_w^* e^{i\sigma_w t}k^* x + \textit{O}(k^{*2}).
\end{equation}
At $\textit{O}(k^{*0})$, the wavy velocity field is indeed the same as the vertical velocity along the foil due to a heaving motion. At $\textit{O}(k^{*0})$, the boundary-value problem for a non-heaving foil in a wavy flow is then the same as that for a heaving foil in a uniform flow. As a result, the pressure distribution along the foil will be the same, leading to equal coefficients of lift and moment. 

The passive pitching motion is given by
\begin{equation}
    \label{eq:long2}
    \theta_{0w} = \frac{\upi i C(\textrm{i} \sigma_w) - \upi\sigma_w}{-\frac{16}{3}R\sigma_w^2 + 4K -2a_{0w} - a_{1w}}V_w^*
\end{equation}
at this order, which is the same as that due to a heaving motion with heaving speed $V_w^*$, but without the inertial forcing proportional to $R$. Given our interest in swimming, we consider the limit $R \rightarrow 0$ in what follows. 

Some of the other quantities of interest differ from those produced by a heaving foil. For a foil passively pitching in a wavy flow, the mean power $C_P$ is zero for any value of $k^*$. The average rate of energy $C_E$ imparted to the fluid is also different, which follows from the fact that $C_E = -C_T$ and the average thrust $C_T$ is the same as that produced by a heaving foil. This follows from the pressure distribution being the same and the slope being the same (since the passive pitching motion is the same). 

At $\textit{O}(k^*)$, the wavy velocity field can be written as 
\begin{equation}
  v_w(x,t) = \left(i V_w^* - k^* V_w^* - \frac{k^* V_w^*}{i \sigma_w} \right) e^{i \sigma_w t} + \frac{k^* V_w^*}{i \sigma_w} e^{i \sigma_w t} + k^* V_w^* e^{i \sigma_w t} (1 + x).
\end{equation}
This takes the same form as the vertical velocity along the foil due to a combined heaving (first term) and pitching (second and third terms) motion. Since we have not considered active pitching motion in the present work, we make no further comments.

\subsection{Short-wave limit}
\label{sec:short}

As $k^* \rightarrow \infty$, the wavelength of the wavy velocity becomes much shorter than the chord. On physical grounds, we expect the wavy velocity to have no effect since any pressure induced by one part of the wavy velocity will be negated by an adjacent part of the wave. In the limit $k^* \rightarrow \infty$, we find that $\theta_w = 0$, and the wavy velocity makes no contribution to $C_L$, $C_M$, $C_P$, $C_E$, or $C_T$. In this limit, the problem is equivalent to setting $V_w = 0$, effectively reducing to a plate with a torsional spring immersed in a uniform flow, which was analyzed by \cite{moore2014analytical}. 

\end{appen}

\bibliographystyle{jfm}
\bibliography{jfm}

@techreport{garrick1936propulsion,
  title={Propulsion of a flapping and oscillating airfoil},
  institution={NACA},
  author={Garrick, I. E.},
  year={1936},
  number={567}
}

@article{moore2014analytical,
  title={Analytical results on the role of flexibility in flapping propulsion},
  author={Moore, M. N. J.},
  journal={Journal of Fluid Mechanics},
  volume={757},
  pages={599--612},
  year={2014},
  publisher={Cambridge University Press}
}

@article{wu1961swimming,
  title={Swimming of a waving plate},
  author={Wu, T. Y.-T.},
  journal={Journal of Fluid Mechanics},
  volume={10},
  number={3},
  pages={321--344},
  year={1961},
  publisher={Cambridge University Press}
}

@article{wu1972extraction,
  title={Extraction of flow energy by a wing oscillating in waves},
  author={Wu, T. Y.-T.},
  journal={Journal of Ship Research},
  volume={16},
  number={01},
  pages={66--78},
  year={1972}
}

@article{smits2019undulatory,
  title={Undulatory and oscillatory swimming},
  author={Smits, A. J.},
  journal={Journal of Fluid Mechanics},
  volume={874},
  pages={P1},
  year={2019},
  publisher={Cambridge University Press}
}

@article{liao2007review,
  title={A review of fish swimming mechanics and behaviour in altered flows},
  author={Liao, J. C.},
  journal={Philosophical Transactions of the Royal Society B: Biological Sciences},
  volume={362},
  number={1487},
  pages={1973--1993},
  year={2007},
  publisher={The Royal Society London}
}

@techreport{theodorsen1935general,
  title={General theory of aerodynamic instability and the mechanism of flutter},
  institution={NACA},
  author={Theodorsen, T.},
  year={1935},
  number={496}
}

@article{von1938airfoil,
  title={Airfoil theory for non-uniform motion},
  author={von K\'arm\'an, T. H. and Sears, W. R.},
  journal={Journal of the Aeronautical Sciences},
  volume={5},
  number={10},
  pages={379--390},
  year={1938}
}

@article{zhang2017footprints,
  title={Footprints of a flapping wing},
  author={Zhang, J.},
  journal={Journal of Fluid Mechanics},
  volume={818},
  pages={1--4},
  year={2017},
  publisher={Cambridge University Press}
}

@article{spagnolie2010surprising,
  title={Surprising behaviors in flapping locomotion with passive pitching},
  author={Spagnolie, S. E. and Moret, L. and Shelley, M. J. and Zhang, J.},
  journal={Physics of Fluids},
  volume={22},
  number={4},
  pages={041903},
  year={2010},
  publisher={American Institute of Physics}
}

@article{moore2015torsional,
  title={Torsional spring is the optimal flexibility arrangement for thrust production of a flapping wing},
  author={Moore, M. N. J.},
  journal={Physics of Fluids},
  volume={27},
  number={9},
  pages={091701},
  year={2015},
  publisher={AIP Publishing LLC}
}

@article{dewey2013scaling,
  title={Scaling laws for the thrust production of flexible pitching panels},
  author={Dewey, P. A. and Boschitsch, B. M. and Moored, K. W. and Stone, H. A. and Smits, A. J.},
  journal={Journal of Fluid Mechanics},
  volume={732},
  pages={29--46},
  year={2013},
  publisher={Cambridge University Press}
}

@article{alben2008optimal,
  title={Optimal flexibility of a flapping appendage in an inviscid fluid},
  author={Alben, S.},
  journal={Journal of Fluid Mechanics},
  volume={614},
  pages={355--380},
  year={2008},
  publisher={Cambridge University Press}
}

@article{floryan2018clarifying,
  title={Clarifying the relationship between efficiency and resonance for flexible inertial swimmers},
  author={Floryan, D. and Rowley, C. W.},
  journal={Journal of Fluid Mechanics},
  volume={853},
  pages={271--300},
  year={2018},
  publisher={Cambridge University Press}
}

@article{baddoo2023generalization,
  title={Generalization of waving-plate theory to multiple interacting swimmers},
  author={Baddoo, P. J. and Moore, N. J. and Oza, A. U. and Crowdy, D. G.},
  journal={Communications on Pure and Applied Mathematics},
  volume={76},
  number={12},
  pages={3811--3851},
  year={2023},
  publisher={Wiley Online Library}
}

@article{goldstein1976complete,
  title={A complete second-order theory for the unsteady flow about an airfoil due to a periodic gust},
  author={Goldstein, M. E. and Atassi, H.},
  journal={Journal of Fluid Mechanics},
  volume={74},
  number={4},
  pages={741--765},
  year={1976},
  publisher={Cambridge University Press}
}

@article{massaro2015effect,
  title={The effect of three-dimensionality on the aerodynamic admittance of thin sections in free stream turbulence},
  author={Massaro, M. and Graham, J. M. R.},
  journal={Journal of Fluids and Structures},
  volume={57},
  pages={81--90},
  year={2015},
  publisher={Elsevier}
}

@article{lysak2013prediction,
  title={Prediction of high frequency gust response with airfoil thickness effects},
  author={Lysak, P. D. and Capone, D. E. and Jonson, M. L.},
  journal={Journal of Fluids and Structures},
  volume={39},
  pages={258--274},
  year={2013},
  publisher={Elsevier}
}

@article{taha2019viscous,
  title={Viscous extension of potential-flow unsteady aerodynamics: the lift frequency response problem},
  author={Taha, H. and Rezaei, A. S.},
  journal={Journal of Fluid Mechanics},
  volume={868},
  pages={141--175},
  year={2019},
  publisher={Cambridge University Press}
}

@article{catlett2020unsteady,
  title={Unsteady response of airfoils due to small-scale pitching motion with considerations for foil thickness and wake motion},
  author={Catlett, M. R. and Anderson, J. M. and Badrya, C. and Baeder, J. D.},
  journal={Journal of Fluids and Structures},
  volume={94},
  pages={102889},
  year={2020},
  publisher={Elsevier}
}

@article{whittlesey2010fish,
  title={Fish schooling as a basis for vertical axis wind turbine farm design},
  author={Whittlesey, R. W. and Liska, S. and Dabiri, J. O.},
  journal={Bioinspiration \& biomimetics},
  volume={5},
  number={3},
  pages={035005},
  year={2010},
  publisher={IOP Publishing}
}

@article{verma2018efficient,
  title={Efficient collective swimming by harnessing vortices through deep reinforcement learning},
  author={Verma, S. and Novati, G. and Koumoutsakos, P.},
  journal={Proceedings of the National Academy of Sciences},
  volume={115},
  number={23},
  pages={5849--5854},
  year={2018},
  publisher={National Acad Sciences}
}

@article{zhu2021numerical,
  title={A numerical study of fish adaption behaviors in complex environments with a deep reinforcement learning and immersed boundary--lattice Boltzmann method},
  author={Zhu, Y. and Tian, F.-B. and Young, J. and Liao, J. C. and Lai, J. C. S.},
  journal={Scientific Reports},
  volume={11},
  number={1},
  pages={1691},
  year={2021},
  publisher={Nature Publishing Group UK London}
}

@article{wu2011fish,
  title={Fish swimming and bird/insect flight},
  author={Wu, T. Y.-T.},
  journal={Annual Review of Fluid Mechanics},
  volume={43},
  pages={25--58},
  year={2011},
  publisher={Annual Reviews}
}

@article{floryan2018efficient,
  title={Efficient cruising for swimming and flying animals is dictated by fluid drag},
  author={Floryan, D. and Van Buren, T. and Smits, A. J.},
  journal={Proceedings of the National Academy of Sciences},
  volume={115},
  number={32},
  pages={8116--8118},
  year={2018},
  publisher={National Acad Sciences}
}

@article{taylor2003flying,
  title={Flying and swimming animals cruise at a {S}trouhal number tuned for high power efficiency},
  author={Taylor, G. K. and Nudds, R. L. and Thomas, A. L. R.},
  journal={Nature},
  volume={425},
  number={6959},
  pages={707--711},
  year={2003},
  publisher={Nature Publishing Group UK London}
}

@article{gazzola2014scaling,
  title={Scaling macroscopic aquatic locomotion},
  author={Gazzola, M. and Argentina, M. and Mahadevan, L.},
  journal={Nature Physics},
  volume={10},
  number={10},
  pages={758--761},
  year={2014},
  publisher={Nature Publishing Group UK London}
}

@article{triantafyllou1991wake,
  title={Wake mechanics for thrust generation in oscillating foils},
  author={Triantafyllou, M. S. and Triantafyllou, G. S. and Gopalkrishnan, R.},
  journal={Physics of Fluids A: Fluid Dynamics},
  volume={3},
  number={12},
  pages={2835--2837},
  year={1991},
  publisher={American Institute of Physics}
}

@article{goza2020connections,
  title={Connections between resonance and nonlinearity in swimming performance of a flexible heaving plate},
  author={Goza, A. and Floryan, D. and Rowley, C.},
  journal={Journal of Fluid Mechanics},
  volume={888},
  pages={A30},
  year={2020},
  publisher={Cambridge University Press}
}

@article{michelin2009resonance,
  title={Resonance and propulsion performance of a heaving flexible wing},
  author={Michelin, S. and Llewellyn Smith, S. G.},
  journal={Physics of Fluids},
  volume={21},
  number={7},
  year={2009},
  publisher={AIP Publishing}
}

@article{paraz2016thrust,
  title={Thrust generation by a heaving flexible foil: Resonance, nonlinearities, and optimality},
  author={Paraz, F. and Schouveiler, L. and Eloy, C.},
  journal={Physics of Fluids},
  volume={28},
  number={1},
  year={2016},
  publisher={AIP Publishing}
}

@article{quinn2014scaling,
  title={Scaling the propulsive performance of heaving flexible panels},
  author={Quinn, D. B. and Lauder, G. V. and Smits, A. J.},
  journal={Journal of Fluid Mechanics},
  volume={738},
  pages={250--267},
  year={2014},
  publisher={Cambridge University Press}
}

@article{heathcote2008effect,
  title={Effect of spanwise flexibility on flapping wing propulsion},
  author={Heathcote, S. and Wang, Z. and Gursul, I.},
  journal={Journal of Fluids and Structures},
  volume={24},
  number={2},
  pages={183--199},
  year={2008},
  publisher={Elsevier}
}

@article{gordnier2013high,
  title={High-fidelity aeroelastic computations of a flapping wing with spanwise flexibility},
  author={Gordnier, R. E. and Chimakurthi, S. K. and Cesnik, C. E. S. and Attar, P. J.},
  journal={Journal of Fluids and Structures},
  volume={40},
  pages={86--104},
  year={2013},
  publisher={Elsevier}
}

@article{liu1997propulsive,
  title={Propulsive performance from oscillating propulsors with spanwise flexibility},
  author={Liu, P. and Bose, N.},
  journal={Proceedings of the Royal Society of London. Series A: Mathematical, Physical and Engineering Sciences},
  volume={453},
  number={1963},
  pages={1763--1770},
  year={1997},
  publisher={The Royal Society}
}

@article{floryan2017scaling,
  title={Scaling the propulsive performance of heaving and pitching foils},
  author={Floryan, D. and Van Buren, T. and Rowley, C. W. and Smits, A. J.},
  journal={Journal of Fluid Mechanics},
  volume={822},
  pages={386--397},
  year={2017},
  publisher={Cambridge University Press}
}

@article{floryan2019large,
  title={Large-amplitude oscillations of foils for efficient propulsion},
  author={Floryan, D. and Van Buren, T. and Smits, A. J.},
  journal={Physical Review Fluids},
  volume={4},
  number={9},
  pages={093102},
  year={2019},
  publisher={APS}
}

@article{van2019scaling,
  title={Scaling and performance of simultaneously heaving and pitching foils},
  author={Van Buren, T. and Floryan, D. and Smits, A. J.},
  journal={AIAA Journal},
  volume={57},
  number={9},
  pages={3666--3677},
  year={2019},
  publisher={American Institute of Aeronautics and Astronautics}
}

@article{floryan2020swimmers,
  title={Swimmers’ wake structures are not reliable indicators of swimming performance},
  author={Floryan, D. and Van Buren, T. and Smits, A. J.},
  journal={Bioinspiration \& Biomimetics},
  volume={15},
  number={2},
  pages={024001},
  year={2020},
  publisher={IOP Publishing}
}

@article{anderson1998oscillating,
  title={Oscillating foils of high propulsive efficiency},
  author={Anderson, J. M. and Streitlien, K. and Barrett, D. S. and Triantafyllou, M. S.},
  journal={Journal of Fluid Mechanics},
  volume={360},
  pages={41--72},
  year={1998},
  publisher={Cambridge University Press}
}

@article{young2007mechanisms,
  title={Mechanisms influencing the efficiency of oscillating airfoil propulsion},
  author={Young, J. and Lai, J. C. S.},
  journal={AIAA Journal},
  volume={45},
  number={7},
  pages={1695--1702},
  year={2007}
}

@article{buchholz2008wake,
  title={The wake structure and thrust performance of a rigid low-aspect-ratio pitching panel},
  author={Buchholz, J. H. J. and Smits, A. J.},
  journal={Journal of Fluid Mechanics},
  volume={603},
  pages={331--365},
  year={2008},
  publisher={Cambridge University Press}
}

@article{chopra1974hydromechanics,
  title={Hydromechanics of lunate-tail swimming propulsion},
  author={Chopra, M. G.},
  journal={Journal of Fluid Mechanics},
  volume={64},
  number={2},
  pages={375--392},
  year={1974},
  publisher={Cambridge University Press}
}

@article{ayancik2019scaling,
  title={Scaling laws for the propulsive performance of three-dimensional pitching propulsors},
  author={Ayancik, F. and Zhong, Q. and Quinn, D. B. and Brandes, A. and Bart-Smith, H. and Moored, K. W.},
  journal={Journal of Fluid Mechanics},
  volume={871},
  pages={1117--1138},
  year={2019},
  publisher={Cambridge University Press}
}

@article{weihs1973hydromechanics,
  title={Hydromechanics of fish schooling},
  author={Weihs, D.},
  journal={Nature},
  volume={241},
  number={5387},
  pages={290--291},
  year={1973},
  publisher={Nature Publishing Group UK London}
}

@incollection{newton2024migration,
title = {Chapter 3: {Migratory flight}},
editor = {Newton, I.},
booktitle = {The Migration Ecology of Birds},
publisher = {Academic Press},
edition = {Second Edition},
address = {Oxford},
pages = {29--50},
year = {2024},
author = {Newton, I.}
}

@article{beal2006passive,
  title={Passive propulsion in vortex wakes},
  author={Beal, D. N. and Hover, F. S. and Triantafyllou, M. S. and Liao, J. C. and Lauder, G. V.},
  journal={Journal of Fluid Mechanics},
  volume={549},
  pages={385--402},
  year={2006},
  publisher={Cambridge University Press}
}

@article{floryan2020distributed,
  title={Distributed flexibility in inertial swimmers},
  author={Floryan, D. and Rowley, C. W.},
  journal={Journal of Fluid Mechanics},
  volume={888},
  pages={A24},
  year={2020},
  publisher={Cambridge University Press}
}

@article{yudin2023propulsive,
  title={Propulsive performance of oscillating plates with time-periodic flexibility},
  author={Yudin, D. and Floryan, D. and Van Buren, T.},
  journal={Journal of Fluid Mechanics},
  volume={959},
  pages={A31},
  year={2023},
  publisher={Cambridge University Press}
}

@techreport{edwards1977unsteady,
  title={Unsteady aerodynamic modeling and active aeroelastic control},
  author={Edwards, J. W.},
  institution={NASA},
  number={NASA-CR-148019},
  year={1977}
}

@article{wu1971hydromechanics,
  title={Hydromechanics of swimming propulsion. Part 1. Swimming of a two-dimensional flexible plate at variable forward speeds in an inviscid fluid},
  author={Wu, T. Y.-T.},
  journal={Journal of Fluid Mechanics},
  volume={46},
  number={2},
  pages={337--355},
  year={1971},
  publisher={Cambridge University Press}
}

@article{fernandez2016linearized,
  title={Linearized propulsion theory of flapping airfoils revisited},
  author={Fernandez-Feria, R.},
  journal={Physical Review Fluids},
  volume={1},
  number={8},
  pages={084502},
  year={2016},
  publisher={APS}
}

@article{gordillo2025note,
  title={A note on the thrust of airfoils},
  author={Gordillo, J. M.},
  journal={Journal of Fluid Mechanics},
  volume={1012},
  pages={A6},
  year={2025},
  publisher={Cambridge University Press}
}

@article{senturk2019reynolds,
  title = {Reynolds Number Scaling of the Propulsive Performance of a Pitching Airfoil},
  author = {Senturk, U. and Smits, A. J.},
  journal = {AIAA Journal},
  volume = {57},
  number = {7},
  pages = {2663--2669},
  year = {2019},
  doi = {10.2514/1.J058371}
}

@article{fernandez2017note,
  title={Note on optimum propulsion of heaving and pitching airfoils from linear potential theory},
  author={Fernandez-Feria, R.},
  journal={Journal of Fluid Mechanics},
  volume={826},
  pages={781--796},
  year={2017},
  publisher={Cambridge University Press}
}

\end{document}